\documentclass[twocolumn,pra,aps,superscriptaddress,longbibliography]{revtex4-2}
\usepackage{amsmath}
\DeclareMathOperator{\sech}{sech}
\usepackage{amssymb}
\usepackage{amsfonts}
\usepackage[dvips]{graphicx}
\usepackage{subfigure}
\usepackage{dcolumn}
\usepackage{txfonts}
\usepackage{bm}
\usepackage{makeidx}
\usepackage{color}
\usepackage{mathtools}
\usepackage{threeparttable}
\usepackage[colorlinks,linkcolor=blue,anchorcolor=blue,citecolor=blue,urlcolor=blue]{hyperref}
\usepackage{lipsum}
\usepackage{braket}
\usepackage{mathrsfs}

\begin{document}

\title{Deterministic Single-Photon Adder and Subtractor}

\author{Si-Qi Jiang}
\affiliation{Center for Quantum Sciences and School of Physics, Northeast Normal University, Changchun 130024, China}

\author{Hai-Jun Xing}
\affiliation{Center for Quantum Sciences and School of Physics, Northeast Normal University, Changchun 130024, China}

\author{Li-Ping Yang$^*$}
\affiliation{Center for Quantum Sciences and School of Physics, Northeast Normal University, Changchun 130024, China}
\email{lipingyang87@gmail.com}

\begin{abstract}
Single-photon addition and subtraction are fundamental operations in quantum information processing. Traditionally, the behavior of a single-photon adder (SPA) and single-photon subtractor (SPS) has been theoretically described using creation and annihilation operators, respectively. However, we demonstrate that this ladder-operator-based description contains significant theoretical flaws. To address these issues, we develop a theoretical framework based on Kraus operators, applicable to both coherent and incoherent SPAs and SPSs. Furthermore, we propose a method for realizing deterministic SPAs and SPSs of a cavity mode using three-level atoms. We analyze the effects of these operations on various quantum states. Additionally, we demonstrate that the use of a control pulse could enhance the performance of SPAs and SPSs, effectively preserving the quantum coherence of the resulting photon state.

\end{abstract}

\maketitle
\section{Introduction}
Single-photon adders (SPAs) and single-photon subtractors (SPSs), which respectively perform the addition or subtraction of a single photon on quantum states, are critical components in quantum information processing. These operations enable the efficient preparation of specialized quantum states, such as optical Schr\"odinger cat states~\cite{Takase2021Generation,Dakna1997Generating,Alexei2006Generating}, entangled states~\cite{Ourjoumtsev2009Preparation,Jeong2014Generation,Morin2014Remote}, and various superposition states~\cite{Neergaard2010Optical,Takahashi2008Generation,Gerrits2010Generation,Zavatta2005Single,Zavatta2007Experimental}. Their precision in quantum state manipulation establishes SPAs and SPSs as indispensable tools not only for quantum state engineering~\cite{Bimbard2010Quantum} and quantum cryptography~\cite{Gisin2002Quantum} but also a building block for photon-based quantum computing~\cite{Menicucci2006Universal,Ferrini2013Compact}.

With the advancement of experimental techniques, SPAs and SPSs have been successfully implemented using various approaches across multiple platforms. An SPS  can be simply achieved using a high-transmittance beam splitter~\cite{Zavatta2008Subtracting,Wenge2004rNon-Gaussian,Neergaard-Nielsen2006Generation,Ourjoumtsev2007Increasing,Parigi2007Probing,Kumar2013Experimental,Fiur2009Engineering}. Once a photon is detected at the reflected port, the single-photon subtraction operation on the incident photon state is successfully performed. An SPA can be realized by exploiting the high nonlinearity of a quantum bit~\cite{Hofheinz2008Generation,Houck2007Generating,Yu-xi2004Generation}, as a single qubit can add at most one photon to the input light at a time. Weakly driven parametric down-conversion and sum-frequency generation processes in nonlinear crystals have also been employed to achieve heralded single-photon addition and subtraction, respectively~\cite{Zavatta2004Quantum,EdoWaks2006Generation,Ra2017Tomography,Brecht2014Demonstration,Manurkar2016Multidimensional,Averchenko2014Nonlinear}. Additionally, the Rydberg blockade effect provides an alternative method for single-photon subtraction. When an alkali atom is excited to a Rydberg state after absorbing a photon from the target light pulse, it changes the energy levels of nearby atoms within its Rydberg radius, thereby preventing the absorption of a second photon~\cite{Tresp2016Absorber,Stiesdal2021Controlled,Murray2018Photon,Honer2011Artificial}. However, realizing coherent and deterministic SPAs and SPSs remains a challenging task in experiments~\cite{Lund2024Subtraction}.


Theoretically, single-photon addition and subtraction operations have been characterized using the creation operator ($\hat{a}^{\dagger}$) and annihilation operator ($\hat{a}$), respectively~\cite{Calsamiglia2001Removal,Pinotsi2008Single,Kim2008Recent,Braun2014Precision,KeyuXia2012Deterministic,Agarwal1991Nonclassical}. Usually, these ladder operators works well for Fock states, i.e., $\hat{a}|n\rangle = \sqrt{n}|n-1\rangle$ and  $\hat{a}^{\dagger}|n\rangle = \sqrt{n+1}|n+1\rangle$ ($n$ is an integer). However, for coherent states, the operator $\hat{a}$ does not alter the state because a coherent state is an eigenstate of the annihilation operator, i.e., $\hat{a}|\alpha\rangle=\alpha|\alpha\rangle$. Moreover, the quantum states resulting from the application of creation or annihilation operators are often not properly normalized. Even if normalization is applied, the resulting state may no longer faithfully represent single-photon addition or subtraction operations. For example, renormalizing the state~$\hat{a}^{\dagger}|\alpha\rangle$ yields~$|\psi\rangle = (|\alpha|^2+1)^{-1/2}\hat{a}^{\dagger}|\alpha\rangle$. However, it can be verified that the average increase in particle number is
\begin{equation}
\langle\psi|\hat{a}^{\dagger}\hat{a}|\psi\rangle - \langle\alpha|\hat{a}^{\dagger}\hat{a}|\alpha\rangle = 1+\frac{|\alpha|^2}{|\alpha|^2+1}.   
\end{equation}
For a large value of $\alpha$, this increase approaches $2$, deviating from the intended single-photon addition. Additionally, the state resulting from the application of ladder operators may, in some cases, not be normalizable, such as $\hat{a}|0\rangle = 0|0\rangle$. Consequently, employing ladder operators to characterize single-photon addition and subtraction presents significant theoretical challenges. This highlights the urgent need for a new theoretical framework for SPAs and SPSs that is universally applicable to all quantum states.

In this work, we first establish the theoretical descriptions of both an ideal coherent and an incoherent SPA using Kraus operators. We also propose a deterministic SPA for a cavity photon mode using a three-level atom. We evaluate the performance of our SPA with three commonly used types of quantum states: superpositions of Fock states, coherent states, and squeezed vacuum states. We numerically simulate the changes in the average photon number, the quantum fluctuations of the quadratures, and the $Q$-function for these three types of states. We show that in the steady state, our proposed SPA deterministically increases the mean photon number of the cavity mode by $1$ for all three types of quantum states. To further enhance the performance of the SPA, we propose adding control pulses to suppress the decoherence of the cavity field induced by the spontaneous decay of the atom. By comparing the performance with that of the ideal SPA, we demonstrate that the quantum coherence of the resulting photon state could be well preserved. Finally, we apply a similar theoretical scheme and numerical analysis to the SPS. 

The paper is organized as follows. In Sec.\ref{sec:adder}, we present a theoretical framework for the SPA and evaluate the performance of our proposed deterministic SPA on three types of quantum states. In Sec.\ref{sec:subtractor}, we apply the similar theoretical approach to the SPS. Section~\ref{sec:conclusion} concludes with a brief summary.

\section{Single-photon adder}
\label{sec:adder} 
Without loss of generality, we focus on single-photon addition and subtraction in a single cavity mode in this work. However, our results could also be applied to quasi-single-frequency long pulses~\cite{Lund2024Subtraction}, by introducing the photon-wavepacket creation operator~\cite{loudon2000quantum}. An ideal coherent SPA for a single mode can be characterized by a Kraus operator 
\begin{equation}
\hat{A}^{\dagger} =\sum^{\infty}_{n=0}\hat{A}^{\dagger}_n=\sum_{n=0}^\infty |n+1\rangle\langle n|,    
\end{equation}
with $\hat{A}\hat{A}^\dagger=\hat{I}_c$ ($\hat{I}_c$
is the identity matrix of the cavity mode). The steady-state density matrix of the cavity after the action of a coherent SPA is given by
\begin{equation}
\rho_{c,ss}^{\rm coh}=\hat{A}^{\dagger}\rho_c(0)\hat{A}, \label{eq:rho_coh}
\end{equation}
where $\rho_c(0)$ is the initial density matrix of the cavity. However, due to the decoherence of the system, the SPA usually functions more like an incoherent SPA. If the cavity is initially in a superposition of Fock states $\left|\psi_{c}\right\rangle=\sum_{n=0}^{\infty}C_{n}\left|n\right\rangle$, each state $|n\rangle$ will correspond to a quantum channel $\hat{A}^{\dagger}_n$, which evolves independently over time as shown in the following. After tracing out additional degrees of freedom of the devices, the quantum coherence of the cavity, characterized by the off-diagonal elements of its density matrix, vanishes in the steady state. Under the action of such an incoherent SPA, the density matrix of the cavity eventually evolves into a mixed state.
\begin{equation}
 \rho_{c,ss}^{\rm incoh}=\sum_{n=0} \hat{A}^\dagger_n\rho_c(0) \hat{A}_n  = \sum_{n=0}^{\infty}P_{n}| n+1\rangle\langle n+1|,  
\end{equation}
with the populations $P_n = |C_n|^2$.

\begin{figure}
\includegraphics[width=7cm]{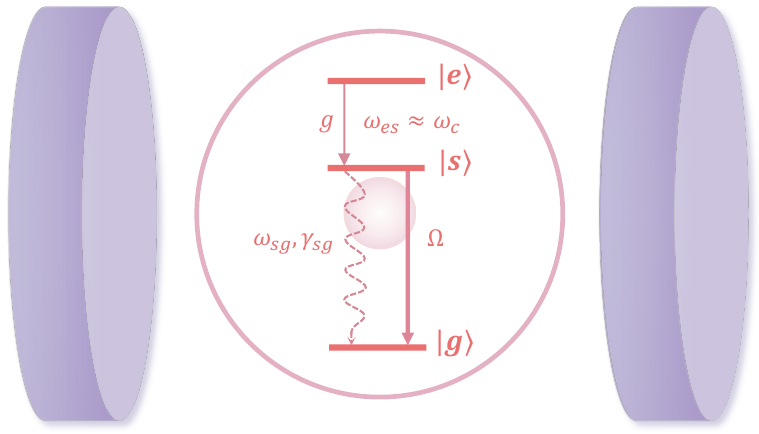}
\centering
\caption{\label{fig:1} Schematic of the single-photon adder (SPA), which consists of a $\Xi$-structured three-level atom placed in a cavity. The $|e\rangle\leftrightarrow |s\rangle$ transition with an interaction strength $g$ is resonant with the microwave cavity mode ($\omega_{es} = \omega_c$ and $\omega_{ij} \equiv \omega_i-\omega_j$). The decay rate from the excited state $|s\rangle$ to the ground state $|g\rangle$ is $\gamma_{sg}$. A control pulse with frequency $\omega_d$ and strength $\Omega$ is applied to stimulate the $|s\rangle\rightarrow|g\rangle$ transition and enhance the performance of the SPA.}
\end{figure}

In this section, we propose a deterministic SPA which consists of a $\Xi$-structured three-level atom coupled to a single-mode cavity field, as shown in Fig.~\ref{fig:1}. The three-level atomic structure includes two excited states, $|e\rangle$ and $|s\rangle$, and one ground state $|g\rangle$. The $|e\rangle\leftrightarrow |s\rangle$ transition is resonant with the microwave cavity mode with an interaction strength $g$. Initially, the atom is in the excited state $|e\rangle$, and it transitions to the state $|s\rangle$ while emitting a microwave photon into the cavity. Finally, due to spontaneous emission, the atom decays to the ground state 
$|g\rangle$, emitting a non-resonant optical photon into free space. This process deterministically adds a photon to the cavity, achieving single-photon addition.

\subsection{Model Hamiltonian}
The Hamiltonian of the whole system contains four parts $\hat{H}=\hat{H}_c+\hat{H}_a + \hat{H}_{\rm int}+\hat{H}_{\rm ctr}$. The cavity field is described by a single-mode harmonic oscillator $\hat{H}_{c}=\hbar\omega_{c}\hat{a}^{\dagger}\hat{a}$ with frequency $\omega_{c}$. The Hamiltonian of three-level atom is given by $\hat{H}_{a}=\hbar\omega_{eg}|e\rangle\langle e| +\hbar\omega_{sg}| s\rangle\langle s|$, where the frequency $\omega_{eg}$ ($\omega_{sg}$) corresponds to the energy difference between state $|e\rangle$ ($|s\rangle$) and state $|g\rangle$. Under the rotating wave approximation, the interaction between the atom and the cavity field is of the Jaynes-Cumming form, 
\begin{equation}
\hat{H}_{\rm int}=\hbar g(\hat{a}\hat{\sigma}_{se}^{\dagger}+\hat{a}^{\dagger}\hat{\sigma}_{se}), \label{eq:Hint}
\end{equation}
where $\hat{\sigma}_{se}=|s\rangle\langle e|$ and $g$ represents the atom-cavity coupling strength. Carefully selecting the atomic states, we can realize the conditions $\omega_c=\omega_{es}=\omega_{eg}-\omega_{sg}$  and $|\omega_{sg}-\omega_c|\gg g$  required by our SPA~\cite{Gleyzes2007jumps}. To improve the performance of the adder as shown in the following, we could add a controlling pulse on the atom
\begin{equation}
 \hat{H}_{\rm ctr} = \hbar\Omega(t)(e^{i\omega_d t}\hat{\sigma}_{gs}+e^{-i\omega_d t}\hat{\sigma}^{\dagger}_{gs}), \label{eq:Hctr}  
\end{equation}
with driving frequency $\omega_d$ and time-varying strength of the control pulse $\Omega (t)$. 

The dynamical evolution of the system can be described by the quantum master equation 
\begin{equation}
\dot{\rho}= -i [\hat{H},\rho]/\hbar + \mathscr{L}\rho, \label{eq:mastereq}   
\end{equation}
where $\mathscr{L\rho}=(\gamma_{sg}/2)(2\hat{\sigma}_{gs}\rho\hat{\sigma}_{gs}^{\dagger}-\{ \sigma_{gs}^{\dagger}\sigma_{gs},\rho\})$ represents the decoherence of the atom with the spontaneous emission rate $\gamma_{sg}$ of the atom from state $|s\rangle$ to state $|g\rangle$. In our SPA model, we have neglected the spontaneous decay of the atom from state $|e\rangle$ to state $|s\rangle$ as well as the cavity field leakage. In experiments~\cite{Gleyzes2007jumps}, the damping time of the microwave cavity could be of scale $\tau_{c}\sim 0.1s$ long. In such a good cavity, the decoherence of the atom corresponding to the resonant transition $|e\rangle\leftrightarrow |s\rangle$ could be greatly suppressed reaching an atomic lifetime of $\tau_{e}\sim 30ms$~\cite{Raimond2001Manipulating}. These two decoherence times are much longer than the typical lifetime (tens of nanoseconds) of an optical transition of natural atoms. Thus, our proposed SPA could be realized in experiments.

\begin{figure}
\includegraphics[width=8cm]{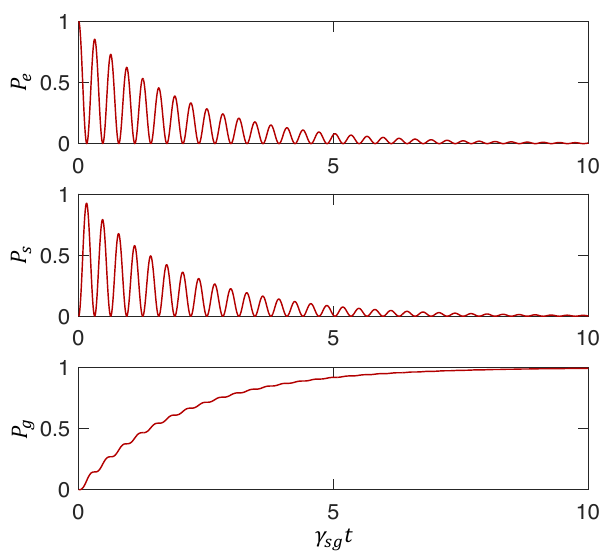}
\centering
\caption{\label{fig:2}The time-varying populations of the three atomic states are shown in the three panels. Rabi oscillations occur between the states $|n,e\rangle$ and $|n+1,s\rangle$ accompanied by the decay of $|n+1,s\rangle$ to $|n+1,g\rangle$. A photon is deterministically added to the microwave cavity in the steady state.}
\end{figure}

\begin{figure}
\includegraphics[width=8cm]{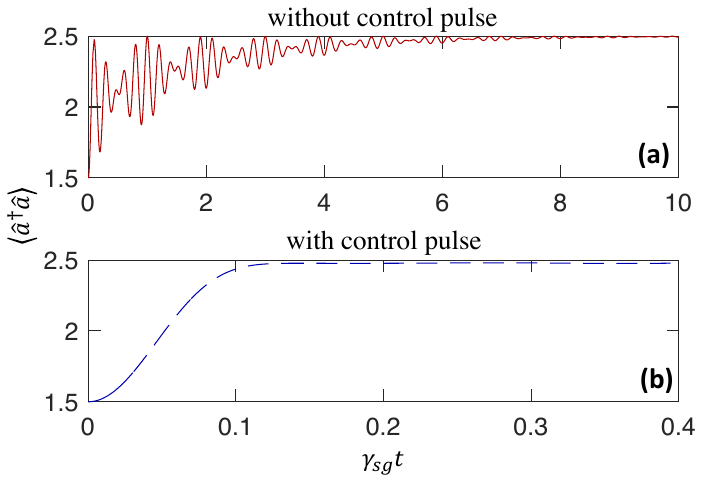}
\centering
\caption{\label{fig:3}Comparison of the mean photon number dynamics in the cavity for the single-photon adder without a control pulse (a) and with a control pulse (b). The cavity is initially in the state $(|1\rangle+|2\rangle)/\sqrt{2}$. The parameters for the optimized square control pulse in Eq.~(\ref{eq:Omega}) are $\Omega/\gamma_{sg} \approx 16$ and $\gamma_{sg}\tau \approx 0.14$. The applied control pulse significantly accelerates the single-photon addition operation.} 
\end{figure}

We first consider a simple case where the cavity is initially in a Fock state $|n\rangle$ and no control pulse is added, i.e., $\Omega (t) = 0$. It can be verified that the three states $\{|n,e\rangle,\ |n+1,s\rangle,\ |n+1,g\rangle\}$ form a closed subspace. The system's dynamics are governed by a set of differential equations:
\begin{align*}
&\dot{\rho}_{n,e;n,e}  =-\sqrt{n+1}ig(\rho_{n+1,s;n,e}-\rho_{n,e;n+1,s})\\
&\dot{\rho}_{n+1,s;n,e}  \!=\!-\sqrt{n+1}ig(\rho_{n,e;n,e}-\rho_{n+1,s;n+1,s})-\frac{\gamma_{sg}}{2}\rho_{n+1,s;n,e}\\
&\dot{\rho}_{n,e;n+1,s}  \!=\!-\sqrt{n+1}ig(\rho_{n+1,s;n+1,s}-\rho_{n,e;n,e})-\frac{\gamma_{sg}}{2}\rho_{n,e;n+1,s}\\
&\dot{\rho}_{n+1,s;n+1,s}  \!=\!-\!\sqrt{n+1}ig(\rho_{n,e;n+1,s}\!-\!\rho_{n+1,s;n,e})\!-\!\gamma_{sg}\rho_{n+1,s;n+1,s}\\
&\dot{\rho}_{n+1,g;n+1,g} \! =\!\gamma_{sg}\rho_{n+1,s;n+1,s}
\end{align*}
with initial conditions $\rho_{ne,ne}(0)=1$, while all other elements of the density matrix are zero. This set of equations can be solved analytically. Here, we only present the numerical results for the populations of the three states in Fig.~\ref{fig:2}. The figure shows that Rabi oscillations occur between the states $|n,e\rangle$ and $|n+1,s\rangle$ accompanied by the decay of $|n+1,s\rangle$ to $|n+1,g\rangle$. In the steady state, a photon is deterministically added to the microwave cavity. In conducting the numerical simulations, we have taken $\gamma=1$ as the unit of the frequency and  $g=10$.

A control pulse in Eq.~(\ref{eq:Hctr}) can be used to improve the performance of the SPA, transforming it from an incoherent SPA to a partially coherent one. In the following sections, we will evaluate the performance of our SPA with three commonly used quantum states: a superposition of a few Fock states, a coherent state, and a squeezed vacuum state.

\subsection{Fock-state case}
In this section, we evaluate the performance of our SPA when acting on a superposition of two Fock states. Without loss of generality, we assume that the cavity is initially in the state $|\psi_{c}\rangle=(|1\rangle+|2\rangle)/\sqrt{2}$ with the mean photon number $\langle\hat{a}^{\dagger}\hat{a}\rangle = 1.5$. In steady state, the mean photon number deterministically increases by $1$ reaching $\langle\hat{a}^{\dagger}\hat{a}\rangle = 2.5$ as shown in Fig.~\ref{fig:3} (a). Here, no control pulse has been applied. For an ideal coherent SPA, the steady state of the cavity should remain a pure state $|\psi_{c,ss}^{\rm coh}\rangle =\hat{A}^{\dagger}|\psi_c\rangle= (|2\rangle+|3\rangle)/\sqrt{2}$. However, we will show that after the operation of the SPA, the cavity loses its quantum coherence and eventually approaches a completely incoherent state in the absence of control pulses.

\begin{figure}
\includegraphics[width=8.5cm]{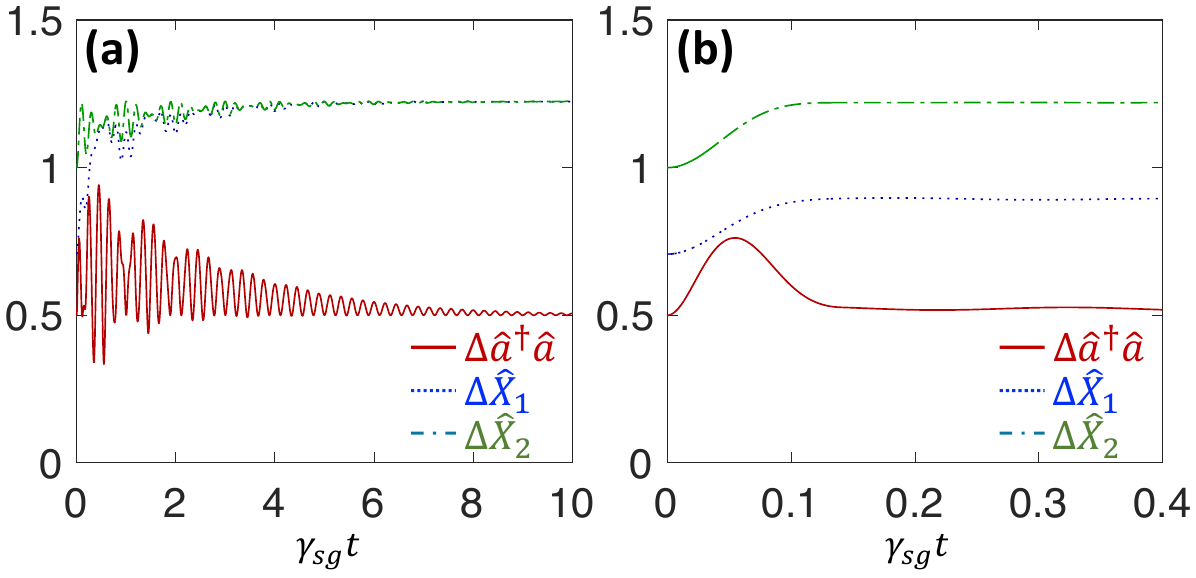}
\centering
\caption{\label{fig:4} Comparison of the performance of the single-photon adder without a control pulse (a) and with a control pulse (b). The three curves in each panel describe the variation of the quantum fluctuations in the photon number and the two quadratures.}
\end{figure}

\begin{figure}
\includegraphics[width=8.5cm]{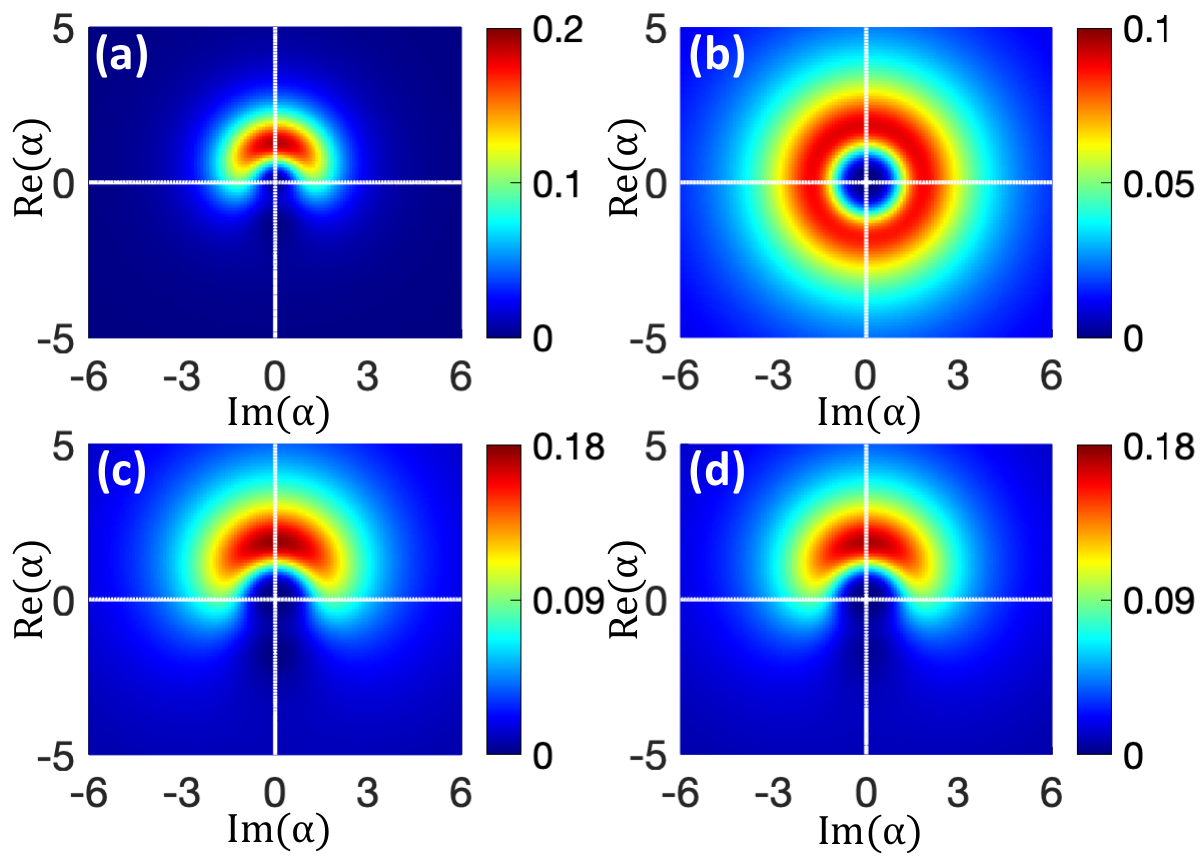}
\centering
\caption{\label{fig:5} Comparison of the performance of incoherent and coherent single-photon adders (SPA) using the $Q$-functions of the cavity mode: (a) the initial state $|\psi_{c}\rangle=(|1\rangle+|2\rangle)/\sqrt{2}$; (b) the steady state $\rho_{c,ss}^{\rm incoh}$ for an incoherent SPA in the absence of a control pulse; (c) the pure state $\hat{A}^{\dagger}|\psi_c\rangle$; (d) the steady state $\rho_{c,ss}$ for SPA in the presence of a square control pulse with an optimized strength $\Omega/\gamma_{sg} \approx 16$ and duration $\gamma_{sg}\tau \approx 0.14$.}
\end{figure}

The off-diagonal elements of the density matrix are usually used to characterize the coherence of a two-level system. For the bosonic cavity mode, we utilize two dimensionless quadratures 
\begin{equation}
\hat{X}_1 = \frac{1}{2}(\hat{a}^{\dagger}+\hat{a}),\ \hat{X}_2 =\frac{1}{2i}(\hat{a}^{\dagger}-\hat{a})\label{eq:X_1} 
\end{equation}
and specifically, their uncertainties in assessing our SPA. As shown in Fig.~\ref{fig:4} (a), the quantum fluctuations of the cavity start from the initial-state values $\Delta \hat{a}^{\dagger}\hat{a}=0.5$, $\Delta \hat{X}_1=0.707$, and $\Delta \hat{X}_2=1$ oscillate with time and finally reach steady values $\Delta \hat{a}^{\dagger}\hat{a}=0.5$, $\Delta \hat{X}_1\approx 1.223$, and $\Delta \hat{X}_2\approx 1.223$. These values are very close to the theoretical values of quantum fluctuations $\Delta \hat{a}^{\dagger}\hat{a}=0.5$, $\Delta \hat{X}_1=1.225$, $\Delta \hat{X}_2=1.225$ when the cavity is in the mixed state $\rho_{c,ss}^{\rm incoh}=(|2\rangle\langle2|+|3\rangle\langle3|)/2$. To dig deeper into the coherence of the cavity, we examine its Husimi $Q$-function $Q(\alpha) = \langle\alpha|\rho_c|\alpha\rangle/\pi$. The $Q$-function of the initial state is a kidney-bean shape in the upper plane as shown in Fig.~\ref{fig:5} (a). Under the operation of the SPA, the $Q$-function spreads out into an almost uniform ring, as shown in panel (b). This confirms the mixed-state nature of the final cavity state.

\begin{figure}
\includegraphics[width=6.5cm]{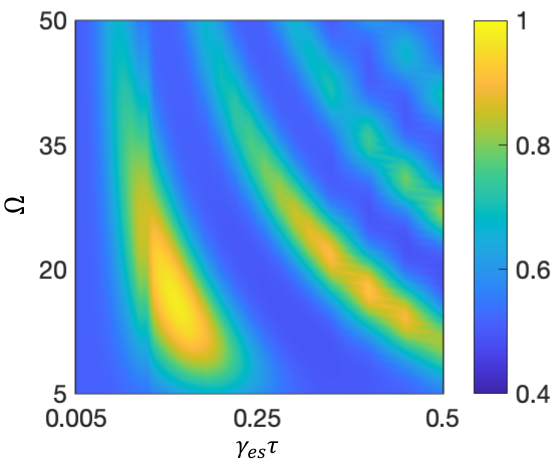}
\centering
\caption{\label{fig:6} Optimization of the control pulse for the single-photon adder operating on a superposition of Fock states. The optimization criterion is the fidelity defined in Eq.~(\ref{eq:Fidelity}), which depends on the pulse strength $\Omega$ and duration $\tau$.}
\end{figure}

To enhance the performance of the SPA, we applied a control pulse resonant with the $|s\rangle\rightarrow |g\rangle$ transition to the atom ($\omega_d = \omega_{sg}$). The decoherence of the cavity primarily arises from the spontaneous decay of the atom from the state~$|s\rangle$. A pulse with a carefully designed shape and duration can effectively suppress this decoherence. For simplicity, we used a rectangular pulse with strength $\Omega$ and duration $\tau$ in our numerical simulations, i.e.,
\begin{equation}
\Omega (t)=\begin{cases}
\Omega, & t\in[0,\tau]\\
0, & {\rm otherwise} \label{eq:Omega} 
\end{cases}.
\end{equation}
To optimize the strength $\Omega$ and pulse length $\tau$, we maximize the fidelity of the steady-state density matrix $\rho_{c,ss}$ of the cavity obtained from the master equation with the density matrix $\rho_{c,ss}^{\rm coh}$ in Eq.~(\ref{eq:rho_coh}) obtained from an ideal SPA
\begin{equation}
 F(\Omega,\tau) =   \left(\rm{Tr}\sqrt{\sqrt{\rho_{c,ss}}\rho_{c,ss}^{\rm coh}\sqrt{\rho_{c,ss}}} \right).\label{eq:Fidelity}
\end{equation}
As shown in Fig.~\ref{fig:6}, the maximum of the fidelity as high as $F\approx 0.977$ locates at $\Omega/\gamma_{sg} \approx 16$ and $\gamma_{sg} \tau \approx 0.14 $.

The coherence of the cavity mode can be preserved quite comprehensively with the application of an optimized control pulse. In Fig.~\ref{fig:3} (b), we demonstrate that the SPA with a control pulse continues to deterministically increase the mean photon number in the cavity by $1$ in a much shorter time.  In this case, the fluctuations of the three quantities in the steady state are given by $\Delta \hat{a}^{\dagger}\hat{a}=0.5$, $\Delta \hat{X}_1=0.895$, and $\Delta \hat{X}_2=1.22$ as shown in Fig.~\ref{fig:4} (b). Compared to the results for an incoherent SPA in Fig.~\ref{fig:4}(a), the fluctuation of the quadrature $\Delta \hat{X}_1$ has been greatly reduced, approaching its theoretical value of $0.866$ of the pure state $|\psi_{c,ss}^{\rm coh}\rangle$. Significant changes in the cavity state can be seen from the $Q$-function of the density matrix in the presence of a control pulse as shown in Fig.~\ref{fig:5} (d). In contrast to panel (b), the bean-shaped structure of the initial state's $Q$-function has been well preserved but with a larger size. The $Q$-function in panel (d) is nearly identical to that obtained from an ideal coherent SPA in panel (c). This indicates that the SPA with an optimized control pulse closely approximates an ideal SPA.

\subsection{Coherent-state case}

\begin{figure}
\includegraphics[width=8cm]{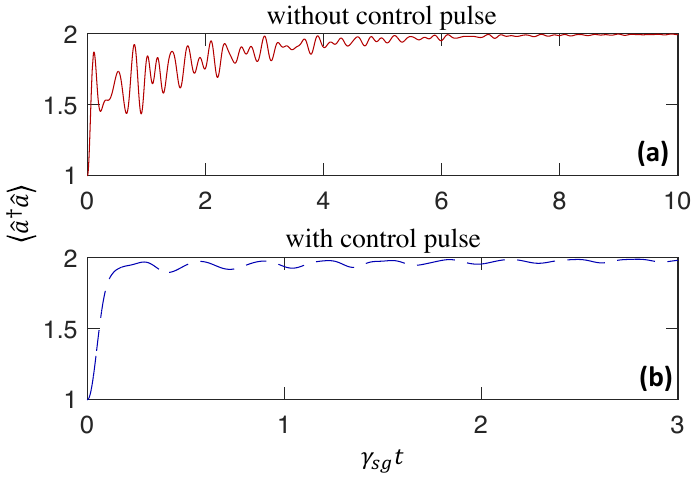}
\centering
\caption{\label{fig:7}Comparison of the mean photon number dynamics in the cavity for the single-photon adder without a control pulse (a) and with a control pulse (b). The cavity is initially in a coherent state $|\psi_{c}\rangle=|\alpha\rangle$ with $\alpha=1$ and mean photon number $\langle\hat{a}^{\dagger}\hat{a}\rangle = |\alpha|^2 =1$. The parameters for the optimized square control pulse in Eq.~(\ref{eq:Omega}) are $\Omega/\gamma_{sg} \approx 15$ and $\gamma_{sg}\tau \approx 0.15$. The applied control pulse significantly accelerates the single-photon addition operation.}
\end{figure}

In this section, we evaluate the performance of our SPA when acting on a coherent state
\begin{equation}
|\psi_{c}\rangle=e^{-\frac{|\alpha|^{2}}{2}}\sum_{n=0}^{\infty}\frac{|\alpha|^{n}}{\sqrt{n!}}|n\rangle
\label{eq:coh}.  
\end{equation}

Without loss of generality, we consider a coherent state with mean photon number $\langle\hat{a}^{\dagger}\hat{a}\rangle = |\alpha|^2 =1$. In numerical simulation, the cutoff dimension of the cavity is safely selected as $N_{\rm cutoff}=15$.  In steady state, the mean photon number deterministically increases by $1$ reaching $\langle\hat{a}^{\dagger}\hat{a}\rangle = 2$ as shown in Fig.~\ref{fig:7} (a). Here, no control pulse has been applied. For an ideal coherent SPA, the steady state of the cavity should remain a pure state 
\begin{equation}
\left|\psi_{c,ss}^{\rm coh}\right\rangle = \hat{A}^{\dagger}|\psi_c\rangle= e^{-|\alpha|^{2}/2}\sum_{n=0}^{\infty}\frac{|\alpha|^{n}}{\sqrt{n!}}|n+1\rangle.    
\end{equation} 
However, we will show that after the operation of the SPA, the cavity loses its quantum coherence and eventually approaches a completely incoherent state.

\begin{figure}
\includegraphics[width=8.5cm]{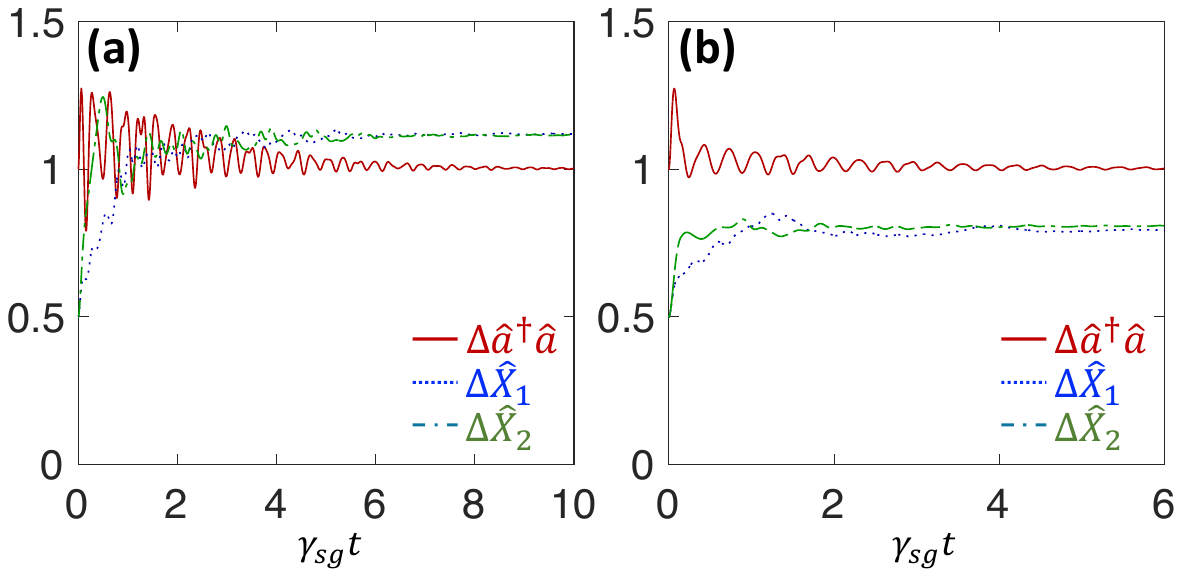}
\centering
\caption{\label{fig:8}Comparison of the performance of the single-photon adder without a control pulse (a) and with a control pulse (b). The three curves in each panel describe the variation of the quantum fluctuations in the photon number and the two quadratures.}
\end{figure}

\begin{figure}
\includegraphics[width=8.5cm]{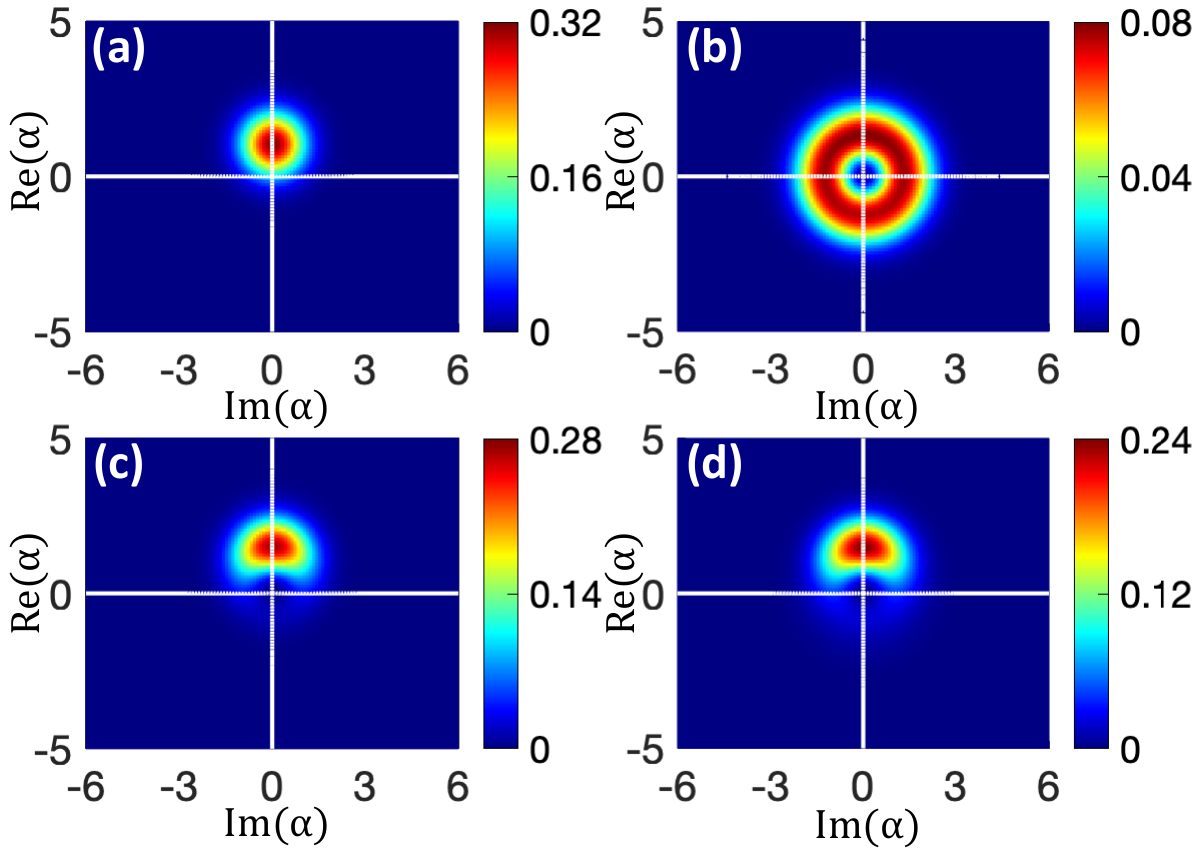}
\centering
\caption{\label{fig:9}Comparison of the performance of incoherent and coherent single-photon adders (SPA) using the $Q$-functions of the cavity mode: (a) the initial coherent state $|\psi_{c}\rangle=|\alpha\rangle$ ($\alpha=1$); (b) the steady state $\rho_{c,ss}^{\rm incoh}$ for an incoherent SPA in the absence of a control pulse; (c) the pure state $\hat{A}^{\dagger}|\psi_c\rangle$; (d) the steady state $\rho_{c,ss}$ for SPA in the presence of a square control pulse with an optimized strength $\Omega/\gamma_{sg} \approx 15$ and duration $\gamma_{sg}\tau \approx 0.15$.}
\end{figure}

\begin{figure}
\includegraphics[width=6.5cm]{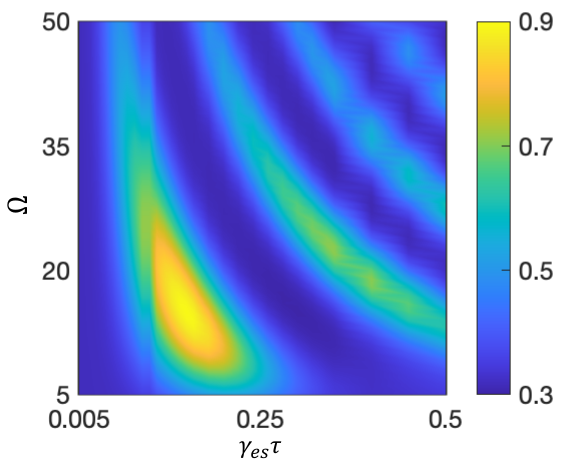}
\centering
\caption{\label{fig:10}Optimization of the control pulse for the single-photon adder operating on a coherent state. The optimization criterion is the fidelity defined in Eq.~(\ref{eq:Fidelity}), which depends on the pulse strength $\Omega$ and duration $\tau$.}
\end{figure}

To characterize the decoherence of the cavity mode induced by the SPA, we simulate the dynamics of the fluctuations in the photon number and the two quadratures as shown in Fig.~\ref{fig:8} (a). The standard deviations of these three operators initially start at 
$\Delta \hat{a}^{\dagger}\hat{a}=1$, $\Delta \hat{X}_1=0.5$, and $\Delta \hat{X}_2=0.5$. They oscillate with time and eventually stabilize at the steady-state values $\Delta \hat{a}^{\dagger}\hat{a}=1$, $\Delta \hat{X}_1\approx 1.119$, and $\Delta \hat{X}_2\approx 1.115$. These steady-state values closely approximate the theoretical quantum fluctuations $\Delta \hat{a}^{\dagger}\hat{a}=1$, $\Delta \hat{X}_1=1.118$, $\Delta \hat{X}_2=1.118$ expected when the cavity is in a mixed state
\begin{equation}
\rho_{c,ss}^{\rm incoh}=e^{-|\alpha|^{2}}\sum_{n=0}^{\infty}\frac{|\alpha|^{2n}}{n!}|n+1\rangle\langle n+1|.  
\end{equation}
To further explore the coherence of the cavity, we examine its Husimi $Q$-function. The $Q$-function of the initial state forms a circular shape in the upper plane, as depicted in Fig.~\ref{fig:9} (a). Upon applying the SPA, the $Q$-function spreads out into a much larger ring center at the origin, as shown in panel (b). This observation confirms the mixed-state nature of the final cavity state. To enhance the performance of the SPA, we applied a rectangular pulse in Eq. (\ref{eq:Omega}) to mitigate the decoherence of the cavity caused by the spontaneous decay of the atom from the state~$|s\rangle$. By varying the strength $\Omega$ and pulse length $\tau$, we optimize the fidelity in Eq.~(\ref{eq:Fidelity}) between the steady-state density matrix $\rho_{c,ss}$ and the density matrix $\rho_{c,ss}^{\rm coh}$ obtained from an ideal coherent SPA. As shown in Fig.~\ref{fig:10}, a maximum fidelity as high as $F\approx 0.896$ can be achieved when  $\Omega/\gamma_{sg} \approx 15$ and $\gamma_{sg} \tau \approx 0.15$.

Under the action of the SPA, the coherence of the cavity mode can be well preserved with the application of an optimized control pulse. In Fig.~\ref{fig:7} (b), we demonstrate that the SPA with a control pulse continues to deterministically increase the mean photon number in the cavity by $1$. In this case, the fluctuations of the three quantities in the steady state are given by $\Delta \hat{a}^{\dagger}\hat{a}=1$, $\Delta \hat{X}_1=0.792$, and $\Delta \hat{X}_2=0.808$ as shown in Fig.~\ref{fig:8} (b). Compared to the results for an incoherent SPA in Fig.~\ref{fig:8} (a), the fluctuations of the quadrature $\Delta \hat{X}_1$ and $\Delta \hat{X}_2$ have been greatly reduced, approaching their theoretical values of $\Delta \hat{X}_1=0.614$ and $\Delta \hat{X}_2=0.71$ of the pure state $|\psi_{c,ss}^{\rm coh}\rangle$.

In contrast to the incoherent SPA case shown in Fig.~\ref{fig:9} (b), significant changes in the cavity state are evident from the $Q$-function of the density matrix in the presence of a control pulse, as shown in panel (d). The circular-shaped structure of the initial state's $Q$-function in panel (a) transforms into a bean-shaped structure of similar size in panel (d). The $Q$-function in panel (d) closely resembles that of an ideal coherent SPA in panel (c), indicating that the SPA with an optimized control pulse closely approximates an ideal SPA.

\subsection{Squeezed-state case}
\begin{figure}
\includegraphics[width=8cm]{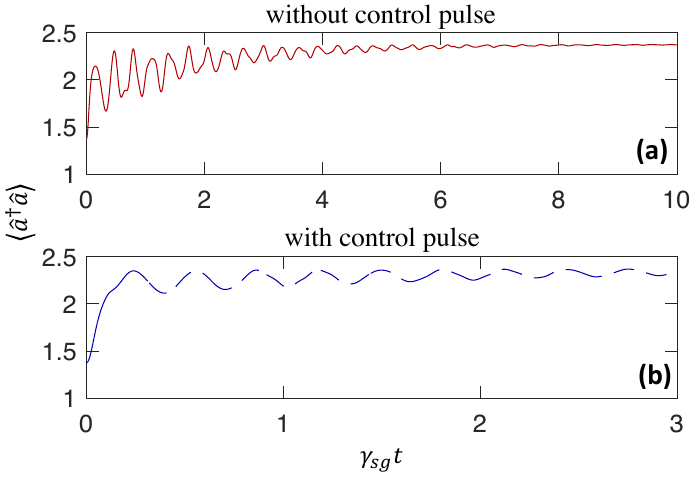}
\centering
\caption{\label{fig:11}Comparison of the mean photon number dynamics in the cavity for the single-photon adder without a control pulse (a) and with a control pulse (b). The cavity is initially in the squeezed vacuum state $|\psi_{c}\rangle$ with squeeze factor $r = 1$ and squeeze angle $\theta = 0$ in Eq.~(\ref{eq:sq}). The parameters for the optimized square control pulse in Eq.~(\ref{eq:Omega}) are $\Omega/\gamma_{sg} \approx 19$ and $\gamma_{sg}\tau \approx 0.12$. The applied control pulse significantly accelerates the single-photon addition operation.}
\end{figure}

In this section, we evaluate the performance of our SPA when acting on a squeezed vacuum state 
\begin{equation}
|\psi_{c}\rangle=(\sech{r})^{\frac{1}{2}}\sum_{n=0}^{\infty}\frac{[(2n)!]^{\frac{1}{2}}}{n!}\left[-\frac{1}{2}e^{i\theta}\tanh{r}\right]^{n}|2n\rangle, \label{eq:sq}  
\end{equation}
where $r$ is the squeeze factor and $\theta$ is the squeeze angle. Without loss of generality, we consider a squeezed vacuum state where $r=1$ and $\theta=0$, leading to a mean photon number of~$\langle\hat{a}^{\dagger}\hat{a}\rangle \approx 1.37$. In numerical simulation, the cutoff dimension of the cavity is safely selected as $N_{\rm cutoff}=25$. In steady state, the mean photon number deterministically increases by $1$ reaching $\langle\hat{a}^{\dagger}\hat{a}\rangle = 2.37$ as shown in Fig.~\ref{fig:11} (a). Here, no control pulse has been applied. For an ideal coherent SPA, the steady state of the cavity should remain a pure state
\begin{equation}
\left|\psi_{c,ss}^{\rm coh}\right\rangle =(\sech{r})^{\frac{1}{2}}\sum_{n=0}^{\infty}\frac{[(2n)!]^{\frac{1}{2}}}{n!}\left[-\frac{1}{2}e^{i\theta}\tanh{r}\right]^{n}|2n+1\rangle. 
\end{equation}
However, we will show that after the operation of the SPA, the cavity also loses its quantum coherence and eventually approaches a completely incoherent state.

\begin{figure}
\includegraphics[width=8.5cm]{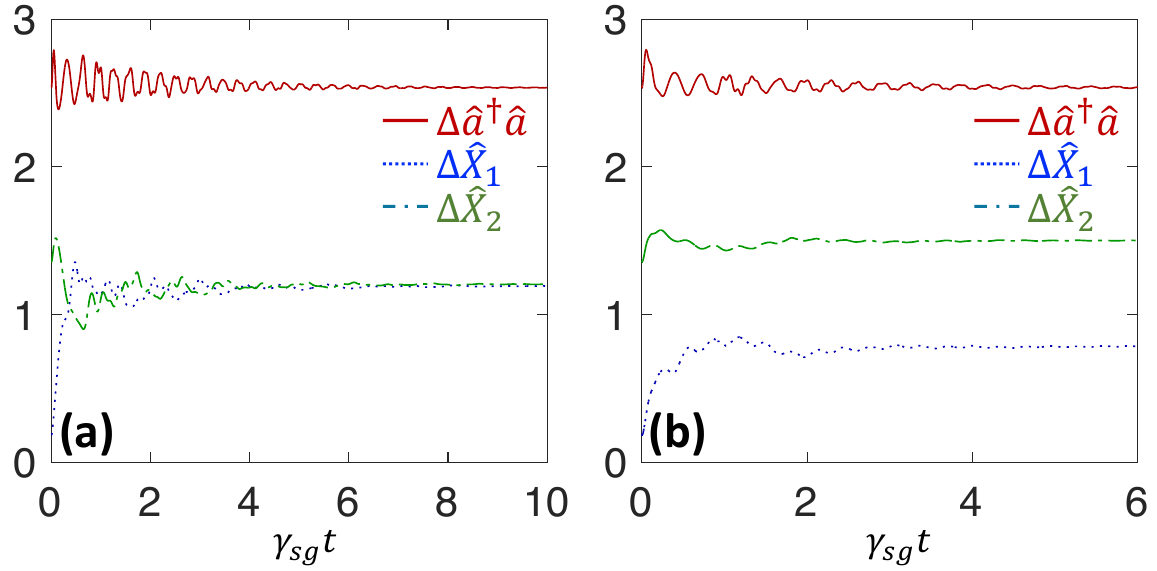}
\centering
\caption{\label{fig:12}Comparison of the performance of the single-photon adder without a control pulse (a) and with a control pulse (b). The three curves in each panel describe the variation of the quantum fluctuations in the photon number and the two quadratures.}
\end{figure}

\begin{figure}
\includegraphics[width=8.5cm]{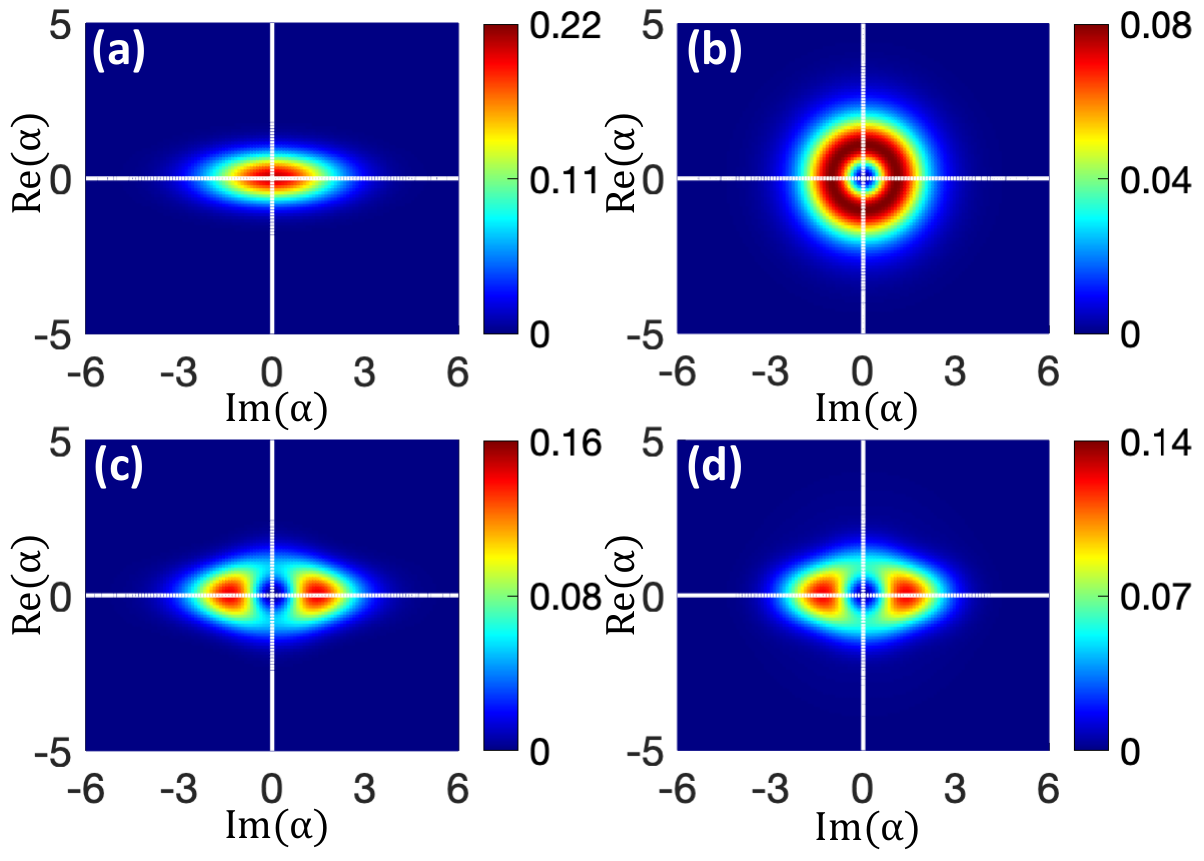}
\centering
\caption{\label{fig:13}Comparison of the performance of incoherent and coherent single-photon adders (SPA) using the $Q$-functions of the cavity mode: (a) the initial squeezed vacuum state $|\psi_{c}\rangle$ with squeeze factor $r=1$ and squeeze angle $\theta=0$; (b) the steady state $\rho_{c,ss}^{\rm incoh}$ for an incoherent SPA in the absence of a control pulse; (c) the pure state $\hat{A}^{\dagger}|\psi_c\rangle$; (d) the steady state $\rho_{c,ss}$ for SPA in the presence of a square control pulse with an optimized strength $\Omega/\gamma_{sg} \approx 19$ and duration $\gamma_{sg}\tau \approx 0.12$.}
\end{figure}

\begin{figure}
\includegraphics[width=6.5cm]{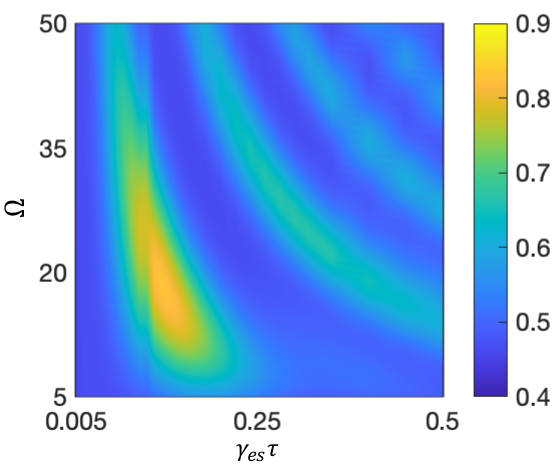}
\centering
\caption{\label{fig:14}Optimization of the control pulse for the single-photon adder operating on a squeezed state. The optimization criterion is the fidelity defined in Eq.~(\ref{eq:Fidelity}), which depends on the pulse strength $\Omega$ and duration $\tau$.}
\end{figure}

To characterize the decoherence of the cavity mode induced by the SPA, we simulate the dynamics of the fluctuations in the photon number and the two quadratures as shown in Fig.~\ref{fig:12} (a). The standard deviations of these three operators initially start at $\Delta \hat{a}^{\dagger}\hat{a}=2.536$, $\Delta \hat{X}_1=0.186$, and $\Delta \hat{X}_2=1.357$. They oscillate with time and eventually stabilize at the steady-state values $\Delta \hat{a}^{\dagger}\hat{a}=2.537$, $\Delta \hat{X}_1\approx 1.192$, and $\Delta \hat{X}_2\approx 1.204$. These steady-state values closely approximate the theoretical the quantum fluctuations $\Delta \hat{a}^{\dagger}\hat{a}=2.521$, $\Delta \hat{X}_1=1.198$, $\Delta \hat{X}_2=1.198$ expected when the cavity is in the mixed state
\begin{equation}
\rho_{c,ss}^{\rm incoh}=\sech r\sum_{n=0}^{\infty}\frac{[(2n)!]}{(n!)^{2}}(-\frac{1}{2}e^{i\theta}\tanh r)^{2n}|2n+1\rangle\langle2n+1|.
\end{equation}
To further explore the coherence of the cavity, we examine its Husimi $Q$-function. The $Q$-function of the initial state is an ellipse shape as shown in Fig.~\ref{fig:13} (a). Under the action of the SPA, the $Q$-function spreads out into an almost uniform ring as shown in panel (b). This confirms the mixed-state nature of the final cavity state. In order to enhance the performance of the SPA, we applied a rectangular pulse with strength $\Omega$ and duration $\tau$ in Eq.~(\ref{eq:Omega}) to suppress the decoherence of the cavity. By varying the strength $\Omega$ and pulse length $\tau$, we optimize the fidelity in Eq.~(\ref{eq:Fidelity}) between the steady-state density matrix $\rho_{c,ss}$ and the density matrix $\rho_{c,ss}^{\rm coh}$ obtained from an ideal coherent SPA. As shown in Fig.~\ref{fig:14}, the maximum fidelity of as high as $F\approx 0.829$ is located at $\Omega/\gamma_{sg} \approx 19$ and $\gamma_{sg} \tau \approx 0.12$.

The coherence of the cavity mode can be preserved quite comprehensively with the application of an optimized control pulse. In Fig.~\ref{fig:11}(b), we demonstrate that the SPA with a control pulse continues to deterministically increase the mean photon number in the cavity by $1$. In this case, the fluctuations of the three quantities in the steady state are given by $\Delta \hat{a}^{\dagger}\hat{a}=2.539$, $\Delta \hat{X}_1=0.781$, and $\Delta \hat{X}_2=1.501$ as shown in Fig.~\ref{fig:12} (b). Compared to the results for an incoherent SPA in Fig.~\ref{fig:12} (a), the fluctuations of the quadrature $\Delta \hat{X}_1$ and $\Delta \hat{X}_2$ have been greatly changed, approaching their theoretical values of $\Delta \hat{X}_1=0.484$ and $\Delta \hat{X}_2=1.624$ of the pure state $|\psi_{c}^{\rm coh}\rangle$.

Compared to the incoherent SPA case as shown in Fig.~\ref{fig:13} (b), significant changes in the cavity state can be seen from the $Q$-function of the density matrix in the presence of a control pulse, as shown in panel (d). The ellipse-shaped structure of the initial state's $Q$-function in panel (a) transforms into a larger size with two peaks in panel (d). The $Q$-function in panel (d) nearly resembles an ideal coherent SPA in panel (c), indicating that the SPA with an optimized control pulse approximates an ideal SPA.

\section{Single-photon subtractor}
\label{sec:subtractor} 
We would expect an ideal coherent SPS characterized by the Kraus operator $\sum_{n=1}^\infty |n-1\rangle\langle n|$. However, this operator does not generate a physical photon state when acting on the vacuum state $|0\rangle$. The correct Kraus operator set for an ideal coherent SPS is given by $\{\hat{A}_0,\hat{A}\}$ with
\begin{equation}
\hat{A}_0=|0\rangle\langle 0|, \ \hat{A}=\sum^{\infty}_{n=1}\hat{A}_n = \sum_{n=1}^\infty |n-1\rangle\langle n|,    
\end{equation} 
which satisfies the completeness $\hat{A}^{\dagger}_0\hat{A}_0+\hat{A}^\dagger \hat{A}=\hat{I}_c$.
It is evident that $\hat{A}_0$, as in practice, illustrates that a subtractor cannot subtract any photons from the vacuum state. After tracing out the atomic degrees of freedom, the steady-state density matrix of the cavity is 
\begin{equation}
\rho_{c,ss}^{\rm coh}=\hat{A}_0\rho_c(0)\hat{A}_0^{\dagger}+\hat{A}\rho_c(0)\hat{A}^{\dagger},\label{eq:rho_coh_sps}
\end{equation}
where $\rho_c(0)$ is the initial density matrix of the cavity. 
We mention that, consistent with the above insight, an ideal coherent SPS can still be realized in two scenarios: (i) The cavity has vanishing initial probability in the vacuum state with $\langle 0 |\rho_c(0)|0\rangle=0$; (ii) One only focuses on the conditional process where a photon has been observed in the environment and the atom is in state $|g\rangle$ in the following SPS.

\begin{figure}
\includegraphics[width=7cm]{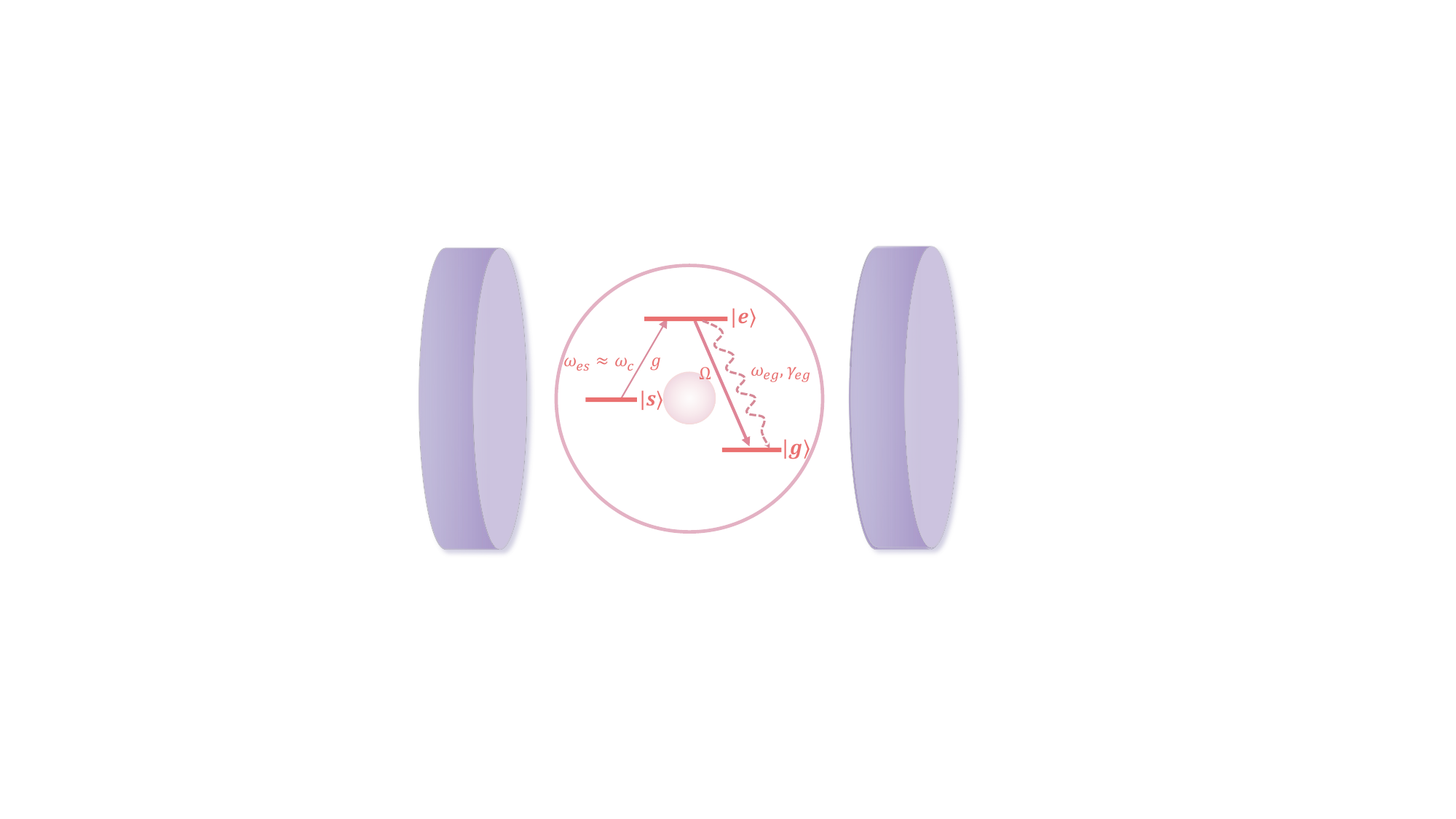}
\caption{\label{fig:15}Schematic of the single-photon subtractor (SPS), which consists of a $\Lambda$-structured three-level atom placed in a cavity. The $|s\rangle\leftrightarrow |e\rangle$ transition with an interaction strength $g$ is resonant with the microwave cavity mode ($\omega_{es} = \omega_c$ and $\omega_{ij} \equiv \omega_i-\omega_j$). The decay rate from the excited state $|e\rangle$ to the ground state $|g\rangle$ is $\gamma_{eg}$. A control pulse with frequency $\omega_d$ and strength $\Omega$ is applied to stimulate the $|e\rangle\rightarrow|g\rangle$ transition and enhance the performance of the SPS.}
\end{figure}

In this section, we introduce an SPS which consists of a $\Lambda$-type three-level atom coupled to a single-mode cavity field, as shown in Fig.~\ref{fig:15}. The three-level atomic structure includes an excited state $|e\rangle$, one ground state $|g\rangle$, and a metastable state $|s\rangle$. Initially, the atom is in the state $|s\rangle$ and transitions to the state $|e\rangle$ while absorbing a microwave photon from the cavity. Finally, due to spontaneous emission, the atom decays to the state 
$|g\rangle$, emitting an optical photon to free space. This process deterministically subtracts a photon from the cavity, achieving single-photon subtraction.

\subsection{Model Hamiltonian}
Similarly to the SPA, the Hamiltonian of the entire system comprises four parts: $\hat{H}=\hat{H}_c+\hat{H}_a + \hat{H}_{\rm int}+\hat{H}_{\rm ctr}$. By carefully selecting the atomic states~\cite{Gleyzes2007jumps}, we can ensure that the cavity mode is resonant with the $|s\rangle\leftrightarrow |e\rangle$ transition ($\omega_c=\omega_{es}$) and significantly detuned from the $|e\rangle\rightarrow |g\rangle$ transition ($|\omega_{eg}-\omega_c|\gg g$) as required by our SPS. The terms $\hat{H}_c$, $\hat{H}_a$, and $\hat{H}_{\rm int}$ are the same as in the SPA. A control pulse nearly resonant with the $|e\rangle\rightarrow |g\rangle$ transition can be applied
\begin{equation}
 \hat{H}_{\rm ctr} = \hbar\Omega(t)(e^{i\omega_d t}\hat{\sigma}_{ge}+e^{-i\omega_d t}\hat{\sigma}^{\dagger}_{ge}), \label{eq:Hctrsub}  
\end{equation}
with driving frequency $\omega_d\approx \omega_{eg}$ and time-varying strength of the control pulse $\Omega(t)$.

The dynamical evolution of the system can be described by the quantum master equation 
\begin{equation}
\dot{\rho}= -i [\hat{H},\rho]/\hbar + \mathscr{L}\rho, \label{eq:mastereqsub}   
\end{equation}
where $\mathscr{L\rho}=(\gamma_{eg}/2)(2\hat{\sigma}_{ge}\rho\hat{\sigma}_{ge}^{\dagger}-\{ \sigma_{ge}^{\dagger}\sigma_{ge},\rho\} )$ represents the decoherence of the atom with the spontaneous emission rate $\gamma_{eg}$ of the atom from state $|e\rangle$ to state $|g\rangle$. In our SPS model, we have neglected the spontaneous decay of the atom from state $|e\rangle$ to state $|s\rangle$ and the leakage of the cavity field. As explained in the SPA section, these approximations can be reasonably achieved in experiments~\cite{Gleyzes2007jumps,Raimond2001Manipulating}.

\begin{figure}
\includegraphics[width=8cm]{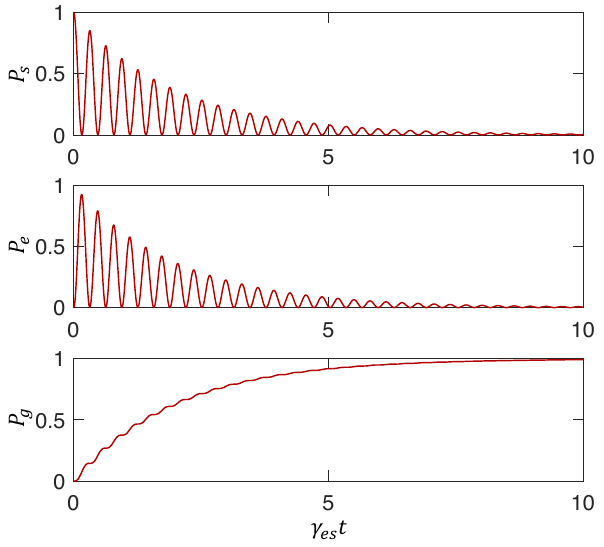}
\caption{\label{fig:16}The time-varying populations of the three atomic states are shown in the three panels. Rabi oscillations occur between the states $|n,s\rangle$ and $|n-1,e\rangle$ accompanied by the decay of $|n-1,e\rangle$ to $|n-1,g\rangle$. A photon is deterministically subtracted from the microwave cavity in the steady state.}
\end{figure}

We first consider a simple case where the cavity is initially in a Fock state $|n\rangle$ ($n\geq 1$) and no control pulse is added, i.e., $\Omega (t) = 0$. It can be verified that the three states $\{|n,s\rangle,\ |n-1,e\rangle,\ |n-1,g\rangle\}$ form a closed subspace. The system's dynamics are governed by a set of differential equations:
\begin{align*}
&\dot{\rho}_{n,s;n,s} =-\sqrt{n}ig(\rho_{n-1,e;n,s}-\rho_{n,s;n-1,e})\\
&\dot{\rho}_{n,s;n-1,e} =-\sqrt{n}ig(\rho_{n-1,e;n-1,e}-\rho_{n,s;n,s})-\frac{\gamma_{eg}}{2}\rho_{n,s;n-1,e}\\
&\dot{\rho}_{n-1,e;n,s} =-\sqrt{n}ig(\rho_{n,s;n,s}-\rho_{n-1,e;n-1,e})-\frac{\gamma_{eg}}{2}\rho_{n-1,e;n,s}\\
&\dot{\rho}_{n-1,e;n-1,e} =-\sqrt{n}ig(\rho_{n,s;n-1,e}-\rho_{n-1,e;n,s})-\gamma_{eg}\rho_{n-1,e;n-1,e}\\
&\dot{\rho}_{n-1,g;n-1,g} =\gamma_{eg}\rho_{n-1,e;n-1,e}
\end{align*}
with initial conditions $\rho_{ns,ns}(0)=1$, while all other elements of the density matrix are zero. This set of equations can be solved analytically. Here, we present only the numerical results for the populations of the three states in Fig.~\ref{fig:16}. The figure shows that Rabi oscillations occur between the states $|n,s\rangle$ and $|n-1,e\rangle$ accompanied by the decay of $|n-1,e\rangle$ to $|n-1,g\rangle$. A photon is deterministically subtracted from the microwave cavity in the steady state. In conducting the numerical simulations, we have taken $\gamma=1$ as the unit of frequency and $g=10$.

Without a control pulse, our SPS functions more like an incoherent SPS. If the cavity is initially in a superposition of Fock states $\left|\psi_{c}\right\rangle=\sum_{n=0}^{\infty}C_{n}\left|n\right\rangle$, each state $|n\rangle$ corresponds to a closed subspace spanned by the three aforementioned states, and these subspaces evolve independently over time. After tracing out the atomic degrees of freedom, the quantum coherence of the cavity, characterized by the off-diagonal elements of its density matrix, vanishes in the steady state. 
Under the action of such an incoherent SPS, the density matrix of the cavity eventually evolves into a mixed state.
\begin{align}
\rho_{c,ss}^{\rm incoh} \! =\!\sum_{n=0}^\infty \hat{A}_n \rho_c(0)\hat{A}_n^\dagger\!=\!P_{0}|0\rangle\langle0|\!+\!\sum_{n=1}^{\infty}P_{n}| n-1\rangle\langle n-1|,
\end{align}
with the populations $P_n =|C_n|^2$.

A control pulse in Eq.~(\ref{eq:Hctrsub}) can be used to improve the performance of the SPS, transforming it from an incoherent SPS to a partially coherent one. In the following sections, we will evaluate the performance of our SPS with three commonly used quantum states: a superposition of a few Fock states, a coherent state, and a squeezed vacuum state.

\subsection{Fock-state case}

\begin{figure}
\includegraphics[width=8cm]{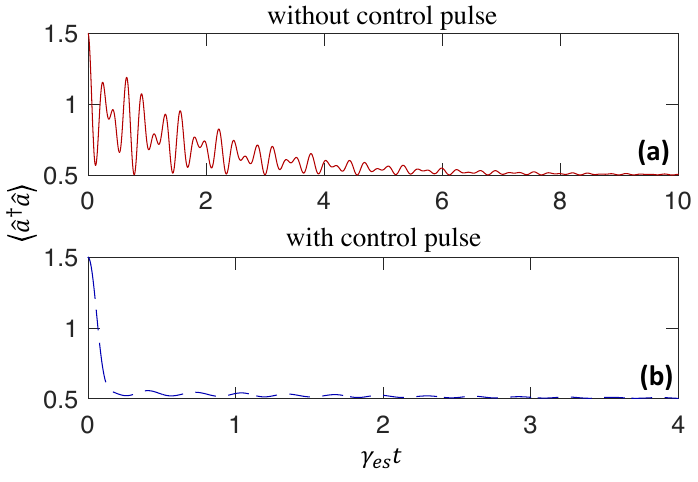}
\centering
\caption{\label{fig:17}Comparison of the mean photon number dynamics in the cavity for the single-photon subtractor without a control pulse (a) and with an optimized control pulse (b). The cavity is initially in the state $(|1\rangle+|2\rangle)/\sqrt{2}$. The parameters for the optimized square control pulse in Eq.~(\ref{eq:Omega}) are $\Omega/\gamma_{sg} \approx 12$ and $\gamma_{sg}\tau \approx 0.18$. The applied control pulse significantly accelerates the single-photon subtraction operation.}
\end{figure}

In this section, we evaluate the performance of our SPS when acting on a superposition of two Fock states. Without loss of generality, we assume that the cavity is initially in the state $|\psi_{c}\rangle=(|1\rangle+|2\rangle)/\sqrt{2}$ with the mean photon number $\langle\hat{a}^{\dagger}\hat{a}\rangle = 1.5$. In steady state, the mean number of photons is deterministically subtracted by $1$ to reach $\langle\hat{a}^{\dagger}\hat{a}\rangle = 0.5$ as shown in Fig.~\ref{fig:17}~(a). Here, no control pulse has been applied. For an ideal coherent SPS, the steady state of the cavity should remain a pure state $|\psi_{c,ss}^{\rm coh}\rangle = \hat{A}|\psi_c\rangle= (|0\rangle+|1\rangle)/\sqrt{2}$. However, we will show that after the operation of the SPS, the cavity loses its quantum coherence and eventually approaches a completely incoherent state in the absence of control pulses.

\begin{figure}
\includegraphics[width=8.5cm]{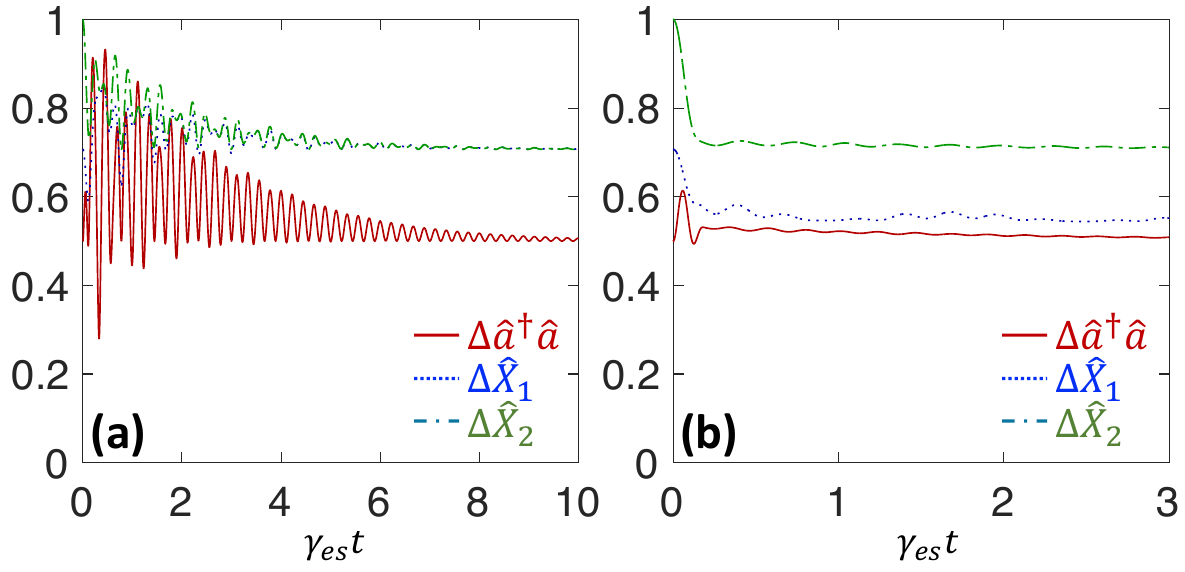}
\centering
\caption{\label{fig:18}Comparison of the performance of the single-photon subtractor without a control pulse (a) and with a control pulse (b). The three curves in each panel describe the variation of the quantum fluctuations in the photon number and the two quadratures.}
\end{figure}

\begin{figure}
\includegraphics[width=8.5cm]{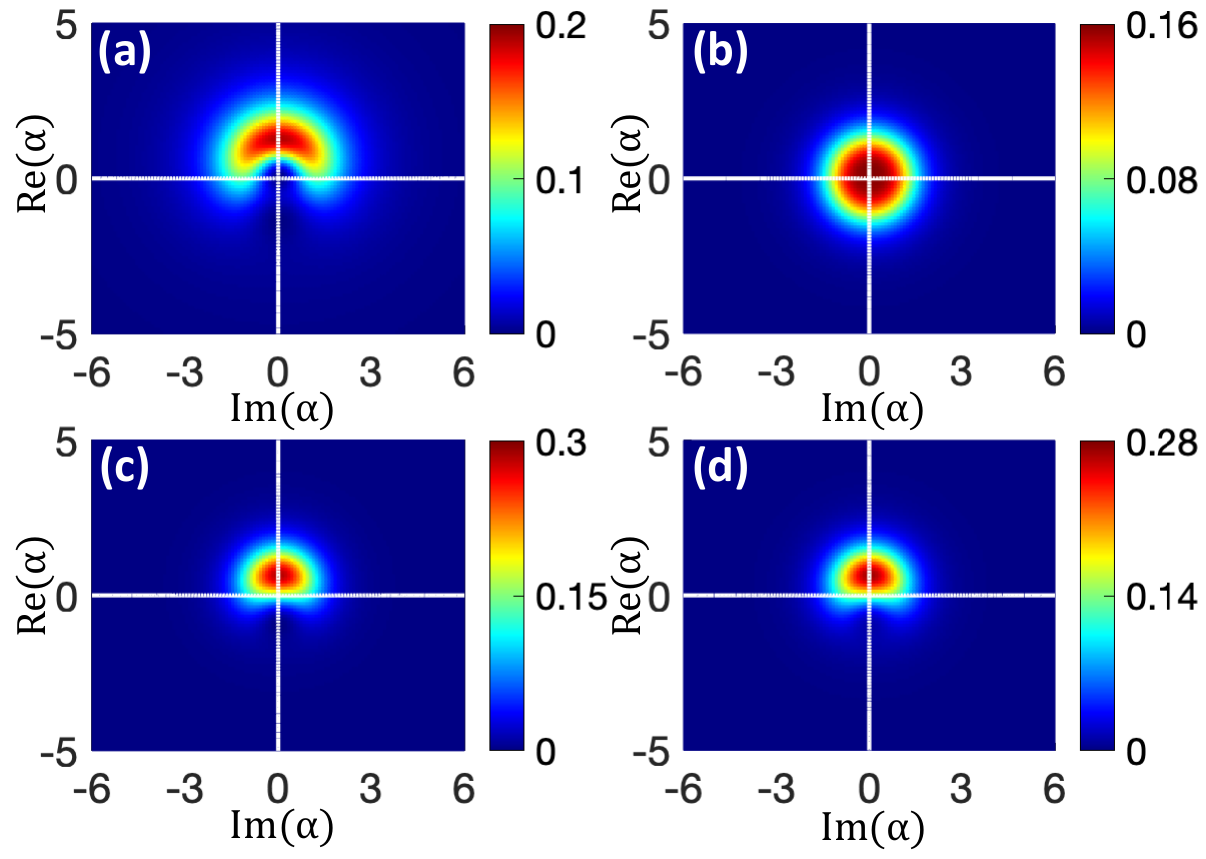}
\centering
\caption{\label{fig:19}Comparison of the performance of incoherent and coherent single-photon subtractors (SPS) using the $Q$-functions of the cavity mode: (a) the initial state $|\psi_{c}\rangle=(|1\rangle+|2\rangle)/\sqrt{2}$; (b) the steady state $\rho_{c,ss}^{\rm incoh}$ for an incoherent SPS in the absence of a control pulse; (c) the pure state $\hat{A}|\psi_c\rangle$; (d) the steady state $\rho_{c,ss}$ for SPS in the presence of a square control pulse with an optimized strength $\Omega/\gamma_{sg} \approx 12$ and duration $\gamma_{sg}\tau \approx 0.18$.}
\end{figure}

We now check the dynamics of the fluctuations in the number of photons and the two quadratures in Eq.~(\ref{eq:X_1}) to evaluate our SPS. As shown in Fig.~\ref{fig:18} (a), the standard deviations of the cavity operators start from the initial-state values $\Delta \hat{a}^{\dagger}\hat{a}=0.5$, $\Delta \hat{X}_1=0.707$, and $\Delta \hat{X}_2=1$ oscillate with time and reach the steady-state values $\Delta \hat{a}^{\dagger}\hat{a}=0.5$, $\Delta \hat{X}_1\approx 0.709$, and $\Delta \hat{X}_2\approx 0.709$. These values are very close to the theoretical values of quantum fluctuations $\Delta \hat{a}^{\dagger}\hat{a}=0.5$, $\Delta \hat{X}_1=0.707$, $\Delta \hat{X}_2=0.707$ when the cavity is in the mixed state $\rho_{c,ss}^{\rm incoh}=(|0\rangle\langle0|+|1\rangle\langle1|)/2$. To dig into the coherence of the cavity, we examine its Husimi $Q$-function. The $Q$-function of the initial state is a kidney-bean shape in the upper plane as shown in Fig.~\ref{fig:19}~(a). Under the operation of the SPS, the $Q$-function function spreads out into a spot centered at the origin as shown in panel (b). This confirms the mixed-state nature of the final cavity state.

\begin{figure}
\includegraphics[width=6.5cm]{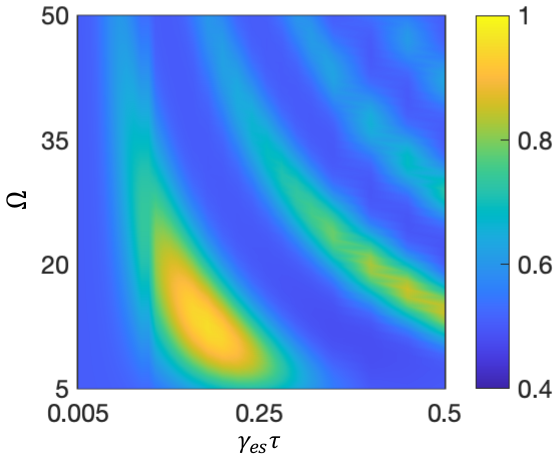}
\centering
\caption{\label{fig:20}Optimization of the control pulse for the single-photon subtractor operating on a superposition of Fock states. The optimization criterion is the fidelity defined in Eq.~(\ref{eq:Fidelity}), which depends on the pulse strength $\Omega$ and duration $\tau$.}
\end{figure}

To enhance the performance of the SPS, we applied a control pulse resonant with the $|e\rangle\rightarrow |g\rangle$ transition to the atom ($\omega_d = \omega_{eg}$). As in SPA, a rectangular pulse in Eq.~(\ref{eq:Omega}) is applied to suppress the decoherence of the cavity, which arises from the spontaneous decay of the atom from the state~$|e\rangle$. we maximize the fidelity in Eq.~(\ref{eq:Fidelity}) of the steady-state density matrix $\rho_{c,ss}$ of the cavity obtained from the master equation (\ref{eq:mastereqsub}) with the density matrix $\rho_{c,ss}^{\rm coh}$ obtained from an ideal SPS in Eq.~(\ref{eq:rho_coh_sps}). As shown in Fig.~\ref{fig:20}, the maximum fidelity as high as $F\approx 0.953$ is located at $\Omega/\gamma_{es} \approx 12$ and $\gamma_{es} \tau \approx 0.18 $. 

The coherence of the cavity mode can be preserved quite comprehensively with the application of an optimized control pulse. In Fig.~\ref{fig:17} (b), we demonstrate that the SPS with a control pulse continues to deterministically subtract the mean photon number in the cavity by $1$ in a much shorter time. In this case, the fluctuations of the three quantities in the steady state are given by $\Delta \hat{a}^{\dagger}\hat{a}=0.5$, $\Delta \hat{X}_1=0.552$, and $\Delta \hat{X}_2=0.711$ as shown in Fig.~\ref{fig:18} (b). Compared to the results for an incoherent SPS in Fig.~\ref{fig:18} (a), the fluctuation of the quadrature $\hat{X}_1$ has been greatly reduced, approaching its theoretical value $\Delta\hat{X}_1=0.5$ in the pure state $\left|\psi_{c,ss}^{\rm coh}\right\rangle$. Significant changes in the cavity state can be seen from the $Q$-function of the density matrix in the presence of a control pulse as shown in Fig.~\ref{fig:19} (d). In contrast to panel (b), the bean-shaped structure of the initial state's $Q$-function has been well preserved but with a smaller size. The $Q$-function in panel (d) is nearly identical to that obtained from an ideal coherent SPS in panel (c). This indicates that the SPS with an optimized control pulse closely approximates an ideal SPS.

\begin{figure}
\includegraphics[width=8cm]{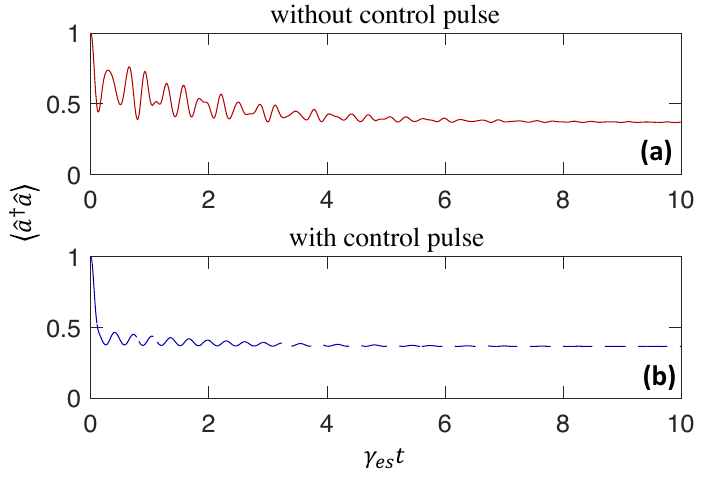}
\centering
\caption{\label{fig:21}Comparison of the mean photon number dynamics in the cavity for the single-photon subtractor without a control pulse (a) and with a control pulse (b). The cavity is initially in the coherent state $|\alpha\rangle$ (\ref{eq:coh}) with mean photon number $\langle\hat{a}^{\dagger}\hat{a}\rangle = |\alpha|^2 =1$. The parameters for the optimized square control pulse in Eq.~(\ref{eq:Omega}) are $\Omega/\gamma_{sg} \approx 16$ and $\gamma_{sg}\tau \approx 0.14$. The applied control pulse significantly accelerates the single-photon subtraction operation.}
\end{figure}

\subsection{Coherent-state case}
In this section, we evaluate the performance of our SPS when acting on a coherent state $|\psi_{c}\rangle$ in Eq.~(\ref{eq:coh}). Without loss of generality, we consider a coherent state with mean photon number $\langle\hat{a}^{\dagger}\hat{a}\rangle = |\alpha|^2 =1$. In numerical simulation, the cutoff dimension of the cavity is safely selected as $N_{\rm cutoff}=15$. In steady state, the mean photon number decreases by $1$ to $\langle\hat{a}^{\dagger}\hat{a}\rangle = 0.37$ as shown in Fig.~\ref{fig:21} (a). Here, no control pulse has been applied. Due to the large vacuum occupation $P_0=0.37$ in the initial coherent state, the reduction in photon number caused by SPS is less than $1$. The vacuum-state component of the coherent state is not coupled to the atom and the SPS remains in the state $|0\rangle\otimes |s\rangle$ during dynamical evolution. For an ideal coherent SPS, the steady state of the whole system will be given by
\begin{equation}
\left|\psi_{c,ss}^{\rm coh}\right\rangle=e^{-\frac{|\alpha|^{2}}{2}}|0\rangle\otimes|s\rangle+e^{-\frac{|\alpha|^{2}}{2}}\sum_{n=1}^{\infty}\frac{|\alpha|^{n}}{\sqrt{n!}}|n-1\rangle\otimes| g\rangle.
\end{equation}
After tracing out the atomic degrees of freedom, the cavity mode will always become a partially coherent state, even for an ideal coherent SPS. However, without a control pulse, the cavity completely loses its quantum coherence and eventually approaches a classical mixed state after the SPS operation.

\begin{figure}
\includegraphics[width=8.5cm]{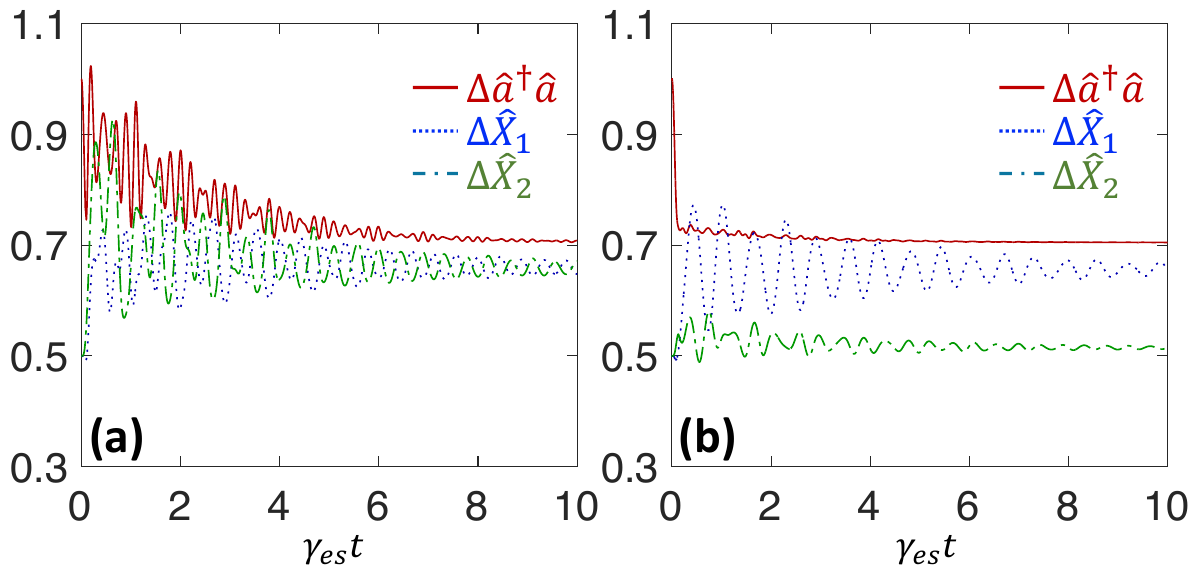}
\centering
\caption{\label{fig:22}Comparison of the performance of the single-photon subtractor without a control pulse (a) and with a control pulse (b). The three curves in each panel describe the variation of the quantum fluctuations in the photon number and the two quadratures.}
\end{figure}

\begin{figure}
\includegraphics[width=8.5cm]{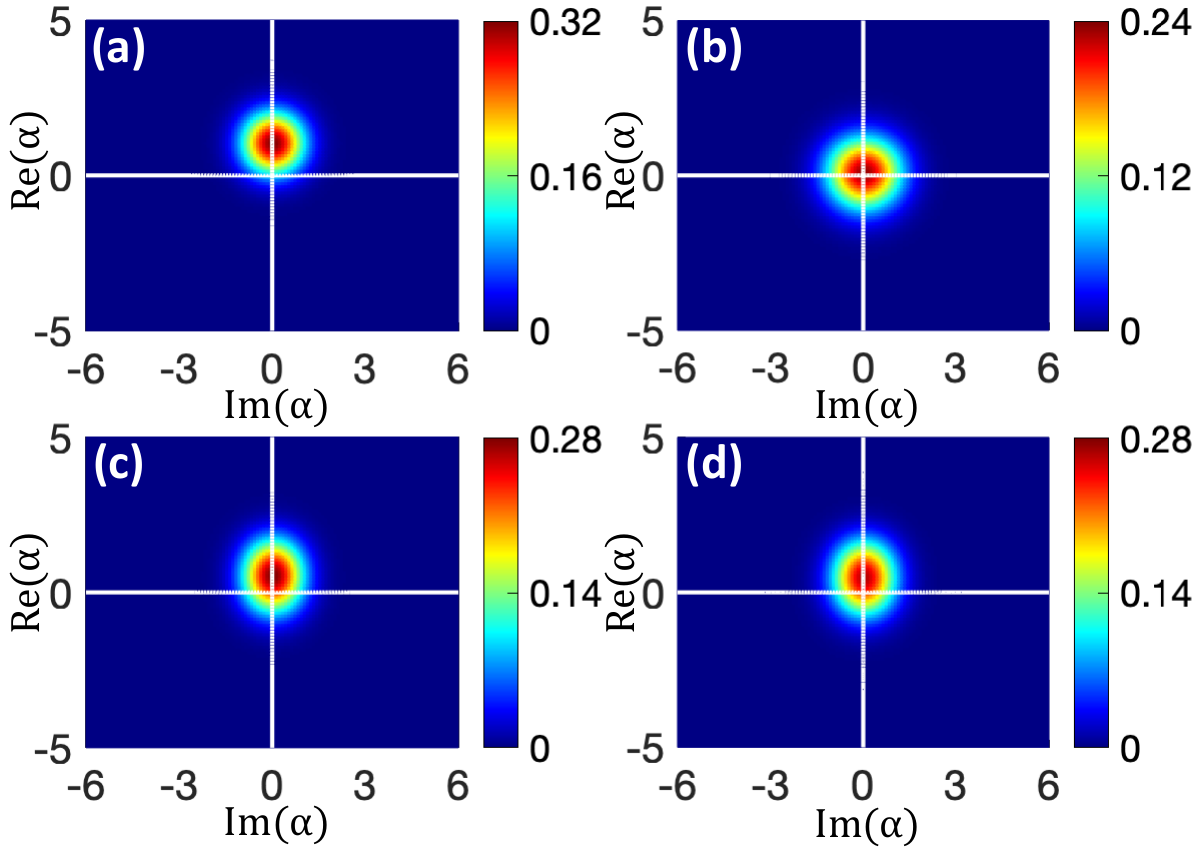}
\centering
\caption{\label{fig:23}Comparison of the performance of incoherent and coherent single-photon subtractors (SPS) using the $Q$-functions of the cavity mode: (a) the initial coherent state $|\alpha\rangle$; (b) the steady state $\rho_{c,ss}^{\rm incoh}$ for an incoherent SPS in the absence of a control pulse; (c) the pure state $\hat{A}|\psi_c\rangle$; (d) the steady state $\rho_{c,ss}$ for SPS in the presence of a square control pulse with an optimized strength $\Omega/\gamma_{sg} \approx 16$ and duration $\gamma_{sg}\tau \approx 0.14$.}
\end{figure}

\begin{figure}
\includegraphics[width=6.5cm]{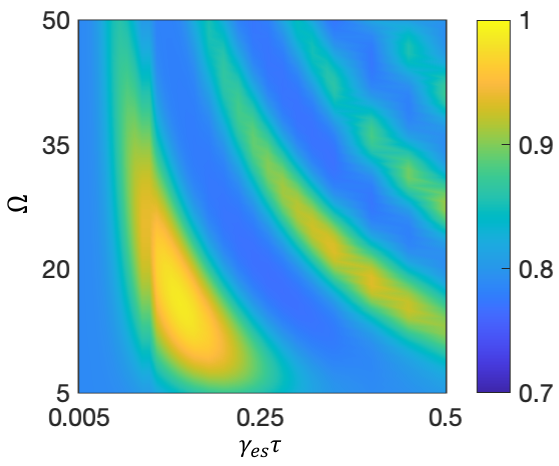}
\centering
\caption{\label{fig:24}Optimization of the control pulse for the single-photon subtractor operating on a coherent state with mean photon number $\langle\hat{a}^{\dagger}\hat{a}\rangle = |\alpha|^2 =1$. The optimization criterion is the fidelity defined in Eq.~(\ref{eq:Fidelity}), which depends on the pulse strength $\Omega$ and duration $\tau$.}
\end{figure}

To assess the performance of our SPS on coherent states, we examine the dynamics of the quantum fluctuations in the photon number and the two quadratures in Eq.~(\ref{eq:X_1}). The standard deviations of these three operators initially start at $\Delta \hat{a}^{\dagger}\hat{a}=1$, $\Delta \hat{X}_1=0.5$, and $\Delta \hat{X}_2=0.5$. They oscillate with time and finally stabilize at the steady values $\Delta \hat{a}^{\dagger}\hat{a}=0.708$, $\Delta \hat{X}_1\approx 0.648$, and $\Delta \hat{X}_2\approx 0.671$ as shown in Fig.~\ref{fig:22} (a). These steady-state values closely approximate the theoretical quantum fluctuations $\Delta \hat{a}^{\dagger}\hat{a}=0.705$, $\Delta \hat{X}_1=0.656$, $\Delta \hat{X}_2=0.656$ expected when the cavity is in a mixed state
\begin{equation}
\rho_{c,ss}^{\rm incoh}=e^{-|\alpha|^{2}}[(1+|\alpha|^{2})|0\rangle\langle0|+\sum_{n=2}^{\infty}\frac{|\alpha|^{2n}}{n!}|n-1\rangle\langle n-1|].
\end{equation}
To further explore the coherence of the cavity, we examine its Husimi $Q$-function. The $Q$-function of the initial state is a circular shape in the upper plane as shown in Fig.~\ref{fig:23} (a). Upon applying the SPS, the $Q$-function gradually becomes a spot centered at the origin as shown in panel (b). This confirms the mixed-state nature of the final cavity state. To enhance the performance of the SPS, we applied a rectangular pulse in Eq.~(\ref{eq:Omega}) to mitigate the decoherence of the cavity. By varying the strength $\Omega$ and pulse length $\tau$, we optimize the fidelity in Eq.~(\ref{eq:Fidelity}) between the steady-state density matrix $\rho_{c,ss}$ and the density matrix $\rho_{c,ss}^{\rm coh}$ obtained from an ideal coherent SPS. As shown in Fig.~\ref{fig:24}, a maximum fidelity as high as $F\approx 0.967$ can be achieved when $\Omega/\gamma_{sg} \approx 16$ and $\gamma_{sg} \tau \approx 0.14$.

Under the action of the SPS, the coherence of the cavity mode can be preserved quite comprehensively with the application of an optimized control pulse. In Fig.~\ref{fig:21} (b), we demonstrate that the SPS with a control pulse continues to decrease the mean photon number in the cavity. In this case, the fluctuations of the three quantities in the steady state are given by $\Delta \hat{a}^{\dagger}\hat{a}=0.705$, $\Delta \hat{X}_1=0.655$, and $\Delta \hat{X}_2=0.515$ as shown in Fig.~\ref{fig:22} (b). Compared to the results for an incoherent SPS in Fig.~\ref{fig:22} (a), the fluctuations of the quadrature $\Delta \hat{X}_1$ and $\Delta \hat{X}_2$ have been greatly reduced, approaching their theoretical values of $\Delta \hat{X}_1=0.637$ and $\Delta \hat{X}_2=0.481$ of the state $\left|\psi_{c,ss}^{\rm coh}\right\rangle$.

In contrast to the incoherent SPS case shown in Fig.~\ref{fig:23} (b), significant changes in the cavity state are evident from the $Q$-function of the density matrix in the presence of a control pulse as shown in panel (d). The circle-shaped structure of the initial state's $Q$-function in panel (a) transforms into an ellipse-shaped structure in panel (d) closely resembles that of an ideal coherent SPS in panel (c), indicating that the SPS with an optimized control pulse closely approximates an ideal SPS.

\begin{figure}
\includegraphics[width=8cm]{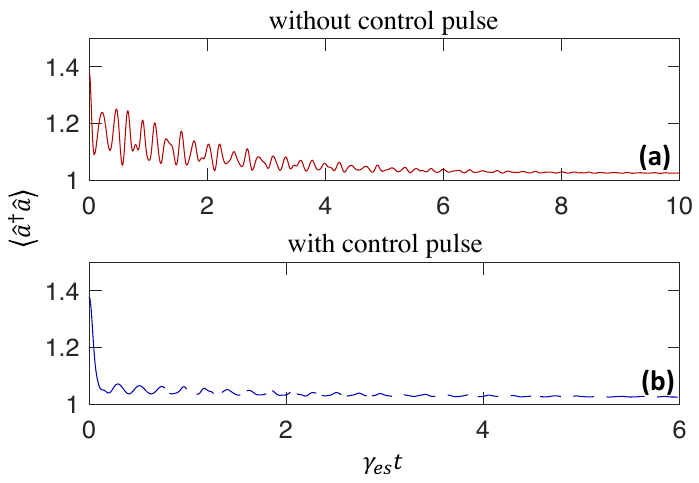}
\centering
\caption{\label{fig:25}Comparison of the mean photon number dynamics in the cavity for the single-photon subtractor without a control pulse (a) and with a control pulse (b). The cavity is initially in the squeezed state (\ref{eq:sq}) with squeeze factor $r = 1$ and squeeze angle $\theta = 0$. The parameters for the optimized square control pulse in Eq.~(\ref{eq:Omega}) are $\Omega/\gamma_{sg} \approx 20$ and $\gamma_{sg}\tau \approx 0.11$. The applied control pulse significantly accelerates the single-photon subtraction operation.}
\end{figure}

\subsection{Squeezed-state case}
In this section, we evaluate the performance of our SPS when acting on a squeezed state $|\psi_{c}\rangle$ in Eq.~(\ref{eq:sq}) with squeeze factor $r=1$ and squeeze angle $\theta=0$. In the absence of a control pulse, the mean photon number decreases from the initial value $1.37$ to $\langle\hat{a}^{\dagger}\hat{a}\rangle = 1.02$ in the steady state, as shown in Fig.~\ref{fig:25} (a). Here, due to high vacuum occupation $P_0=0.65$ in the initial squeezed state, the photon number reduction caused by SPS is significantly less than $1$. For an ideal coherent SPS, the steady state of the whole system will be given by
\begin{align}
\left|\psi_{c,ss}^{\rm coh}\right\rangle = & (\sech{r})^{\frac{1}{2}}\{|0\rangle\otimes|s\rangle\nonumber\\
& +\sum_{n=1}^{\infty}\frac{[(2n)!]^{\frac{1}{2}}}{n!}\left(-\frac{1}{2}\tanh{r}\right)^{n}|2n-1\rangle\otimes|g\rangle\}. \label{eq:psi_coh_sq}
\end{align}

\begin{figure}
\includegraphics[width=8.5cm]{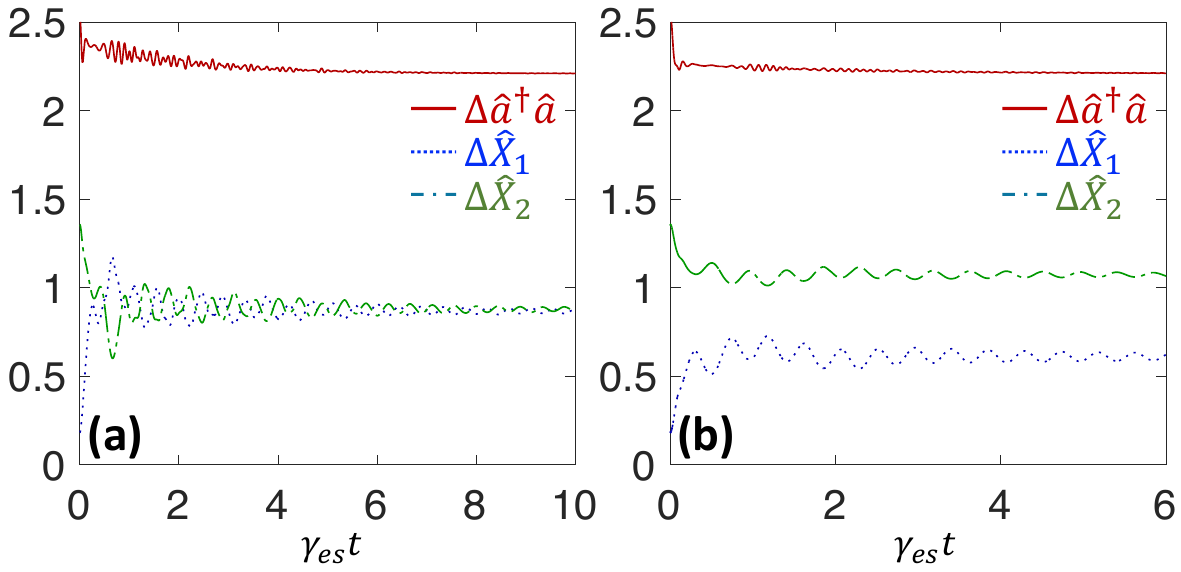}
\centering
\caption{\label{fig:26}Comparison of the performance of the single-photon subtractor without a control pulse (a) and with a control pulse (b). The three curves in each panel describe the variation of the quantum fluctuations in the photon number and the two quadratures.}
\end{figure}

To characterize the decoherence of the cavity mode induced by the SPS, we simulate the dynamics of the fluctuations in the photon number and two quadratures as shown in Fig.~\ref{fig:26} (a). The standard deviations of these three operators initially start at 
$\Delta \hat{a}^{\dagger}\hat{a}=2.536$, $\Delta \hat{X}_1=0.186$, and $\Delta \hat{X}_2=1.357$. They oscillate with time and finally reach the steady values $\Delta \hat{a}^{\dagger}\hat{a}=2.21$, $\Delta \hat{X}_1\approx 0.879$, and $\Delta \hat{X}_2\approx 0.867$. These steady-state values are very close to the theoretical values of the quantum fluctuations $\Delta \hat{a}^{\dagger}\hat{a}=2.209$, $\Delta \hat{X}_1=0.873$, $\Delta \hat{X}_2=0.873$ expected when the cavity is in the mixed state 
\begin{equation}
\rho_{c,ss}^{\rm incoh}\!=\!\sech{r}\left\{\!|0\rangle\langle0|\!+\!\!\sum_{n=1}^{\infty}\!\frac{(2n)!}{(n!)^{2}}\!\left(\frac{\tanh{r}}{2}\right)^{2n}\!|2n\!-\!1\rangle\langle 2n\!-\!1|\!\right\}.
\end{equation}

\begin{figure}
\includegraphics[width=8.5cm]{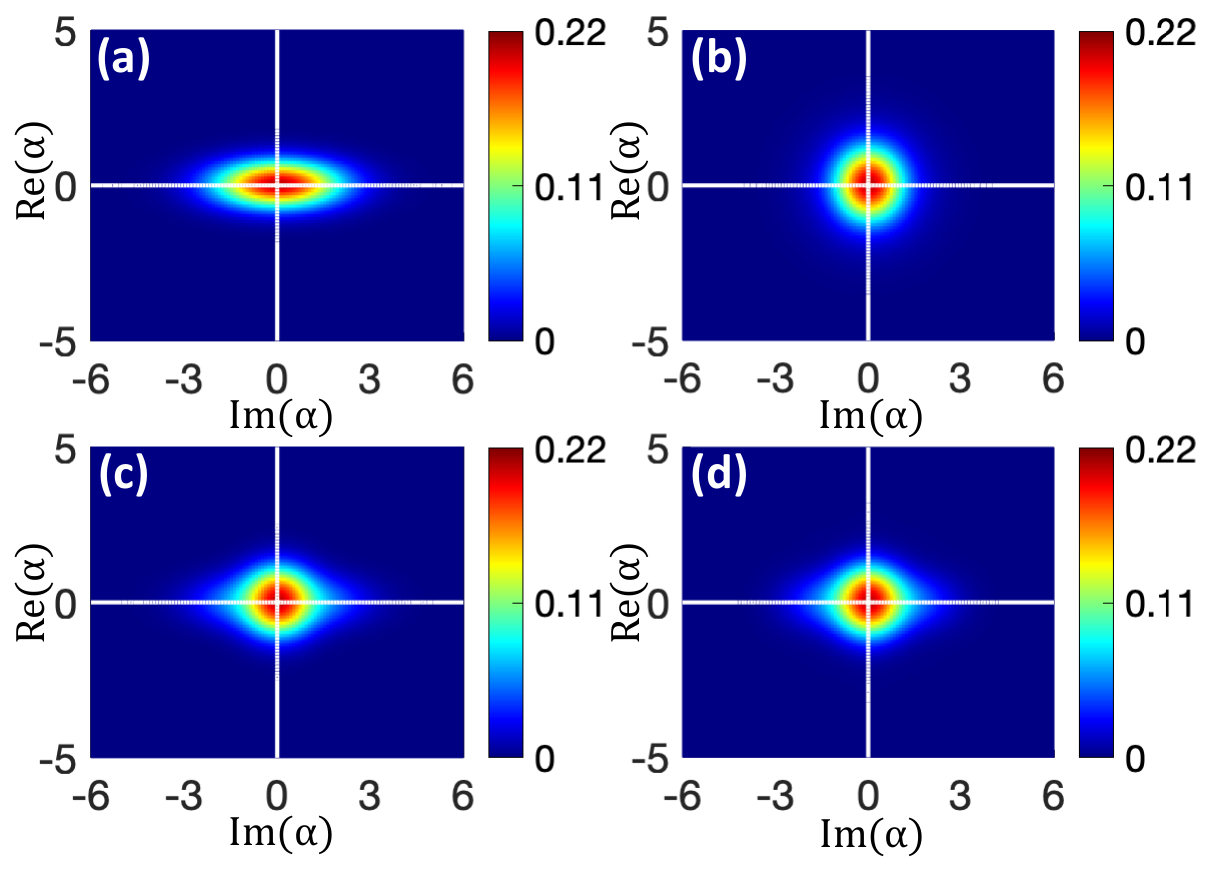}
\centering
\caption{\label{fig:27}Comparison of the performance of incoherent and coherent single-photon subtractors (SPS) using the $Q$-functions of the cavity mode: (a) the initial state $|\psi_{c}\rangle=(\sech{r})^{1/2}\sum_{n=0}^{\infty}([(2n)!]^{1/2}[-\exp{(i\theta)}\tanh{r}/2]^{n}/n!)|2n\rangle$ with squeeze factor $r=1$ and squeeze angle $\theta=0$; (b) the steady state $\rho_{c,ss}^{\rm incoh}$ for an incoherent SPA in the absence of a control pulse; (c) the pure state $\hat{A}{\dagger}|\psi_c\rangle$; (d) the steady state $\rho_{c,ss}$ for SPS in the presence of a square control pulse with an optimized strength $\Omega/\gamma_{sg} \approx 20$ and duration $\gamma_{sg}\tau \approx 0.11$.}
\end{figure}

\begin{figure}
\includegraphics[width=6.5cm]{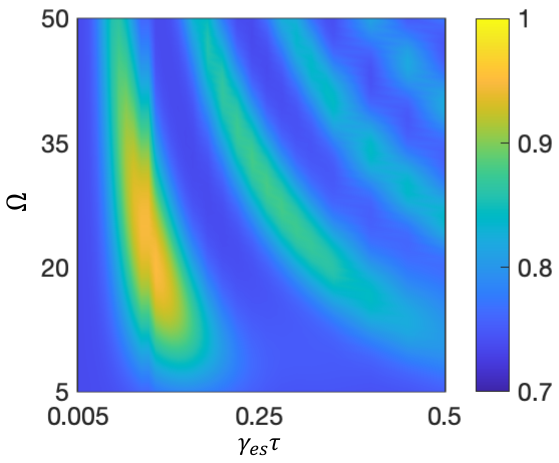}
\centering
\caption{\label{fig:28}Optimization of the control pulse for the single-photon subtractor operating on a squeezed state. The optimization criterion is the fidelity defined in Eq.~(\ref{eq:Fidelity}), which depends on the pulse strength $\Omega$ and duration $\tau$.}
\end{figure}

To further explore the coherence of the cavity, we examine its Husimi $Q$-function. The $Q$-function of the initial state is an ellipse shape as shown in Fig.~\ref{fig:27} (a). Under the operation of the SPS, the $Q$-function gradually becomes a spot centered at the origin as shown in panel (b). This confirms the mixed-state nature of the final cavity state. To enhance the performance of the SPS, we applied a rectangular pulse with strength $\Omega$ and duration $\tau$ in Eq.~(\ref{eq:Omega}) to suppress the decoherence of the cavity. By varying the strength $\omega$ and pulse length $\tau$, we optimize the fidelity in Eq.~(\ref{eq:Fidelity}) between the steady-state density matrix $\rho_{c,ss}$ and the density matrix $\rho_{c,ss}^{coh}$ obtained from an ideal coherent SPS. As shown in Fig.~\ref{fig:28}, an maximum fidelity $F\approx 0.951$ is obtained at $\Omega/\gamma_{sg} \approx 20$ and $\gamma_{sg} \tau \approx 0.11$.

The coherence of the cavity mode can be well preserved with the help of an optimized control pulse. In Fig.~\ref{fig:25} (b), we demonstrate that the SPS with a control pulse still reduces the mean photon number in the cavity by approximately $0.35$. In this case, the fluctuations of the three quantities in the steady state are given by $\Delta \hat{a}^{\dagger}\hat{a}=2.21$, $\Delta \hat{X}_1=0.623$, and $\Delta \hat{X}_2=1.066$ as shown in Fig.~\ref{fig:26} (b). Compared to the results for an incoherent SPS in Fig.~\ref{fig:26} (a), the fluctuations of the quadrature $\Delta \hat{X}_1$ and $\Delta \hat{X}_2$ are closer to the expected results $\Delta \hat{X}_1=0.476$ and $\Delta \hat{X}_2=1.139$ for the pure state in Eq.~(\ref{eq:psi_coh_sq}). In contrast to the incoherent SPS case shown in Fig.~\ref{fig:27} (b), significant changes in the cavity state are evident from the $Q$-function of the density matrix in the presence of a control pulse, as shown in panel (d). The ellipse-shaped structure of the initial state's $Q$-function in panel (a) transforms into a circle-shaped structure in panel (d) closely resembles that of an ideal coherent SPS in panel (c), indicating that the SPS with an optimized control pulse closely approximates an ideal SPS.

\section{Conclusion \label{sec:conclusion}}
In this work, we established a Kraus operator-based theoretical framework to rigorously model SPAs and SPSs, enabling precise characterization of their action on quantum states. We proposed a deterministic SPA design for a single cavity mode utilizing a three-level atom and validated its performance through numerical simulations. For three widely used quantum states, our results demonstrate that the SPA reliably increases the cavity mode’s average photon number by one in the steady state. Furthermore, by implementing tailored control pulses to surpress cavity field decoherence induced by spontaneous atomic decay, we significantly enhance the preservation of quantum coherence, thereby improving the SPA’s operational fidelity. Extending this framework to SPSs, we also developed a unified formalism for single-photon subtraction operations. Comparative analyses between our model and ideal SPA and SPS benchmarks confirm the theoretical validity of our approach. We derive analytical steady-state expressions for quantum states undergoing single-photon addition and subtraction and examine the resulting changes in state properties. These findings may aid in optimizing the experimental implementation of SPAs and SPSs for future quantum technology applications.

\section*{Acknowledgment}
This work is supported by the Innovation Program for Quantum Science and Technology (No.~2023ZD0300700), the National Natural Science Foundation of China (NSFC) (Grant No.~12275048), and the Fundamental Research Funds for the Central Universities (Grant No.~2412023QD007).

\bibliography{main}
\end{document}